\documentclass[aps]{revtex4-1}
\usepackage{amsfonts}
\pdfoutput=1
\usepackage{amsmath}
\usepackage{amssymb}
\usepackage{graphicx}

\input{tcilatex}

\begin{document}

\title{Critical ultrasonic propagation in magnetic fields}
\author{A. Pawlak}
\email{pawlak@amu.edu.pl}
\date{\today }

\begin{abstract}
Effect of an external magnetic field on the critical sound attenuation and
velocity of the longitudinal wave is studied in ferromagnets. We derive a
parametric model that incorporates a crossover from the asymptotic critical
behavior to the Landau-Ginzburg regular behavior far away from the critical
point. The dynamics is based on the time dependent Ginzburg-Landau model
with non conserved order parameter (model A). The variations of the sound
attenuation coefficient and velocity have been obtained for arbitrary values
of the magnetic field and reduced temperature. The scaling functions are
given within the renormalization group formalism at one-loop order. Using
MnP as an example, we show that such parametric crossover model yields an
accurate description of ultrasonic data in a large region of temperatures
and magnetic fields around the critical point.

\begin{description}
%\item[PACS numbers] 05.70.Jk; 64.60.Ht; 75.10; 75.40.Gb
\item[Keywords]critical behavior, ultrasonic attenuation, sound dispersion, Ginzburg-Landau model, RG method, parametric model
\end{description}

\end{abstract}
%Keywords
%{critical behavior, ultrasonic attenuation, sound dispersion, parametric model}
%\pacs{05.70.Jk; 62.65.+k; 64.60.Ht; 75.40.Gb}
\maketitle

\affiliation{ Faculty of Physics, Adam Mickiewicz University,
Uniwersytetu Poznañskiego 2, 61-614 Pozna\'{n}, Poland}

\preprint{APS/123-QED}

% Force line breaks with \\




%%%%%%%%%%%%%%%%%%%%%%%%%%%%%%%%%
\section{Introduction}
\vspace{-5mm}
As the sound wave strongly couples with the order parameter fluctuations the
critical dynamics of sound is a research field where we can test modern
concepts of the phase transition theory such as the universality of critical
exponents, scaling or the crossover to another universality class etc. Sound
attenuation and velocity in magnetic systems have been extensively studied
both experimentally and theoretically \cite{Luthi81}-\cite{Paw11} . Whereas,
the temperature behavior of these quantities has been relatively well
recognized during last decades \cite%
{lut04,Deng87,Paw98,Paw00,Erdem04,PawFech06} using mean-field theories,
scaling theory and renormalization group framework a relatively little
attention has been paid so far to the problem of critical attenuation in the
presence of the ordering magnetic field \cite%
{Kom74,Tach74,Paw09,Paw12,Paw20,Paw22}.

The singular behavior of the sound attenuation coefficient is connected with
very strong fluctuations of the magnetic order parameter near the critical
temperature. These fluctuations give rise to a characteristic attenuation
peak whose position is correlated with a dip in the sound velocity \cite%
{lut04}. In metallic magnets the critical anomalies in the sound attenuation
coefficient are of different types than in magnetic insulators. The
difference in the critical exponents describing these anomalous behaviors
are usually explained by the occurrence of different kinds of magnetoelastic
coupling in the two classes of magnets mentioned \cite{Kaw68}. It was also
shown \cite{Paw98} that a very important role is played by the ratio of the
spin-lattice relaxation time to the characteristic time of spin
fluctuations. It is a crucial parameter determining whether the sound
attenuation coefficient reveals a strong or a weak singularity in a given
material. In this paper we will focus on the class of metallic ferromagnets
which are characterized by strong anomaly of the sound attenuation
coefficient (sometimes called Murata-Iro-Schwabl singularity \cite%
{Mur76,IroSchwabl83}) connected with \ short spin-lattice relaxation times
\cite{Paw98,Paw11}.  In these magnetic materials the maximum in the sound
attenuation coefficient occurs in the ordered phase for vanishing magnetic
field due to the domination of the magnetic analogue of the
Landau-Khalatnikov sound damping near the superfluid transition of liquid $%
^{4}$He \cite{LK54}. Such mechanism can be explained as a   result of
relaxation of   the average order-parameter and should be distinguished from
fluctuation contribution described by  relaxation of a non-conserved
variable which is proportional to the square of the order parameter. \ The
Landau-Khalatnikov term is the only one which contributes to the ultrasound
attenuation in the mean-field theory. However in the scaling region (close
to the critical temperature), as was noted by Dengler and Schwabl \cite%
{Deng87} \ both terms combine just to one single scaling function as there
will be cancelations between the various contributions \cite{HH77}.\ When a
magnetic field is applied the maximum in the sound attenuation coefficient
is shifted towards higher temperatures and can be located even at
temperatures much higher than the Curie temperature \cite{Kom74,Tach74,Paw20}%
.   Another characteristic feature of the \ ultrasonic attenuation in
magnetic field is that the increase in the sound attenuation \ in high
temperature phase is much faster than in the low temperature phase as we
approach the critical point. The steeper rise in the high temperature region
may be explained by the increase of the magnetization induced by the field.
However, finding a single expression to describe the ultrasonic attenuation
in a broad range of temperatures and magnetic fields is a very difficult
problem that has not been fully resolved so far. As was shown recently \
\cite{Paw20} away from the critical point a sixth order term in
Landau-Ginzburg energy may be also of importance.

It is the aim of this paper to describe the magnetic field and temperature
dependence of ultrasonic attenuation   and sound velocity in ferromagnets in
terms of a parametric crossover model \cite{Paw22} which incorporates both
the asymptotic critical behavior near the Curie temperature as well as the
Ginzburg-Landau behavior (with sixth-order term included in the free energy)
as we cross over to large magnetic fields or for the temperatures not very
close to the Curie temperature. We illustrate our findings with ultrasonic
data in manganese phosphide MnP. A good agreement is achieved for a wide
range of temperatures and magnetic fields. Our results successfully explain
the rapid increase in the sound attenuation in the high temperature phase
and the shift of the maximum of attenuation towards the higher temperatures
as the magnetic field increases. The magnetic field as well as frequency
dependence of the hight of this maximum can be understood within this model.

The paper is organized as follows. In Sec. II we define the model and recall
the equivalent functional form  of the equations of motion.   The results of
the dynamic renormalization group for the acoustic self-energy are presented
in Sec. III. We discuss the   Landau-Ginzburg approximation in Sec. IV. Sec.
V presents the details of our parametric crossover model where we have
included the   sixth-order term in Landau-Ginzburg expansion of the free
energy density.   We also discuss the sound attenuation coefficient and
sound dispersion scaling functions in this representation. In Sec VI we
compare the theoretical results found in this work with the experimental
data obtained by Komatsubara at al. \cite{Kom74} and Ishizaki et al. \cite%
{Ishi77} for longitudinal sound wave propagating in MnP.   Finally, the
conclusions are summarized in Sec. VII.

\section{ MODEL}

\subsection{Statics}

We consider Ising-like ($n=1$) continuous order parameter\ \ on a $d$%
-dimensional elastic solid \cite{Paw98}:
\begin{equation}
\mathcal{H}=\mathcal{H}_{\mathrm{OP}}+\mathcal{H}_{\mathrm{el}}+\mathcal{H}_{%
\mathrm{int}},  \label{ham1}
\end{equation}%
where $\mathcal{H}_{\mathrm{OP}}(S)$ is the Ginzburg-Landau part of the free
energy for the one-component order parameter $S(\mathbf{x})$:
\begin{equation}
\mathcal{H}_{\mathrm{OP}}=\int d^{d}\!x\left[ \frac{1}{2}r_{0}S(\mathbf{x}%
)^{2}+\frac{1}{2}(\mathbf{\nabla }S)^{2}+\frac{u}{4}S(\mathbf{x})^{6}+\frac{v%
}{6}S(\mathbf{x})^{4}\right]  \label{hop}
\end{equation}%
and
\begin{equation}
\mathcal{H}_{\mathrm{el}}=\frac{1}{2}\int d^{d}\!x\left\{ B\varepsilon _{ii}(%
\mathbf{x})^{2}+2\mu \left[ \varepsilon _{ij}(\mathbf{x})-\frac{1}{d}\delta
_{ij}\varepsilon _{ll}(\mathbf{x})\right] ^{2}\right\}  \label{hel}
\end{equation}%
is the elastic contribution in the harmonic approximation, with $\varepsilon
_{ij}(\mathbf{x})$ denoting the strain tensor related to the displacement
vector components $u_{i}(\mathbf{x})$ by
\begin{equation*}
\varepsilon _{ij}(\mathbf{x})=\frac{1}{2}(\nabla _{i}u_{j}+\nabla _{j}u_{i}).
\end{equation*}%
The first term in Eq.(\ref{hel}) describes the contribution due to the
volume changes ($\varepsilon _{ii}(\mathbf{x})$ is proportional to the
density fluctuation) and the second term is due to the shear distortions
\cite{bergman,luben}. $B$ and \ \ $\mu $ are the (bare) bulk and shear
modulus (the unitary mass density and $k_{B}T_{C}=1$ have been assumed),
respectively. For simplicity, we have assumed the solid to be isotropic. \
Finally, the interaction Hamiltonian is given by
\begin{equation}
\mathcal{H}_{\mathrm{int}}=f_{0}\int d^{d}\!x\varepsilon _{ii}(\mathbf{x})S(%
\mathbf{x})^{2},  \label{hint}
\end{equation}%
which describes the volume magnetostriction with the coupling constant $%
f_{0} $. In the magnetic part $r_{0}=at$ is linear in the reduced
temperature $t=\frac{T-T_{C}}{T_{C}}$ and the parameters $u$ are $v$ \ are
coupling constants. It is well known that $v$ is irrelevant for the critical
behavior.

A given elastic configuration $\varepsilon _{ij}(\mathbf{x})$ can be
separated \cite{larkin,Paw89} into a homogenous deformation and \ the
constant-volume phonon part:
\begin{equation}
\varepsilon _{ij}(\mathbf{x})=\varepsilon _{ij}^{0}+\frac{1}{2V}\sum_{%
\mathbf{k}\neq 0}[k_{i}u_{j}(\mathbf{k})+k_{_{j}}u_{i}(\mathbf{k})]\exp (i%
\mathbf{k\cdot x}),  \label{lar}
\end{equation}%
where $V$ is the volume of the system at equilibrium. We decompose the
displacement vector $\mathbf{u}(\mathbf{k})$\ \ into longitudinal and
transverse parts defined via
\begin{equation}
\mathbf{u(k)}=\widehat{\mathbf{k}}Q_{\mathrm{L}}(\mathbf{k})+\mathbf{Q}_{%
\mathrm{T}}(\mathbf{k}),  \label{longtr}
\end{equation}%
where $Q_{\mathrm{L}}(\mathbf{k})=\widehat{\mathbf{k}}\cdot \mathbf{u(k)},\
\mathbf{Q}_{\mathrm{T}}(\mathbf{k})=\mathbf{u(k)-}\widehat{\mathbf{k}}Q_{%
\mathrm{L}}(\mathbf{k})$ \ and $\widehat{\mathbf{k}}=\mathbf{k/}\left\vert
\mathbf{k}\right\vert .$ \ Only the longitudinal mode $Q_{\mathrm{L}}$ \ is
coupled, in this model, to the order parameter fluctuations so we integrate
over the transverse modes as well as over homogenous deformations $%
\varepsilon _{ij}^{0}$ in the partition function. This is equivalent to
considering the system under a fixed external pressure \cite{nattermann}. As
a result we get a new Hamiltonian whose elastic part takes a simple form
\begin{equation*}
\mathcal{H}_{\mathrm{el}}=\int \frac{d^{d}\!k}{(2\pi )^{d}}%
k^{2}c_{0}^{2}\left\vert Q_{\mathrm{L}}(\mathbf{k})\right\vert ^{2}
\end{equation*}%
with $c_{0}^{2}=B+2(d-1)\mu /d$ \ as the velocity square of the longitudinal
sound mode of the non-interacting system. The interaction part is
transformed into:%
\begin{equation}
\mathcal{H}_{\text{int}}=f_{0}\int \frac{d^{d}\!k}{(2\pi )^{d}}kQ_{\mathrm{L}%
}(\mathbf{k})S_{-\mathbf{k}}^{2},
\end{equation}%
where $S_{\mathbf{k}}^{2}=\int \frac{d^{d}\!p}{2\pi }S(\mathbf{p})S(\mathbf{k%
}-\mathbf{p})$.

\subsection{Dynamics}

In an isotropic solid, which is considered here for simplicity, the
transverse sound decouple from the order parameter and will be neglected. We
consider only the longitudinal sound and order-parameter modes which are
mutually coupled to each other. On the grounds of the hydrodynamics and
renormalization group arguments, the following system of Langevin equations
\cite{Paw98}, is considered:
\begin{equation}
\dot{S}_{\mathbf{k}}=-\Gamma \frac{\delta \mathcal{H}}{\delta S_{\mathbf{-k}}%
}+\xi _{\mathbf{k}},  \label{rr1}
\end{equation}%
\begin{equation}
\ddot{Q}_{\mathbf{k}}=-\frac{\delta \mathcal{H}}{\delta Q_{\mathbf{-k}}}%
-\Theta k^{2}\dot{Q}_{\mathbf{k}}+\eta _{\mathbf{k}},  \label{rr2}
\end{equation}%
where the index 'longitudinal' for the elastic modes has been omitted. The
Fourier components of the Gaussian white noises $\xi $ nd $\eta $ \ have
variances related to the bare damping terms $\Gamma $and \ $\Theta k^{2}$\ \
through the usual Einstein relations. Here, $\Theta k^{2}$ is responsible
for the noncritical sound dumping and $\Gamma $ is a relaxation coefficient
of the order parameter.

It is convenient to represent the model in terms of the equivalent
functional form \cite{Janss79,Domin77}\ \ with an Onsager-Machlup functional
$\mathcal{J}(S,Q;\widetilde{S},\widetilde{Q}),$ where auxiliary
\textquotedblright response\textquotedblright\ fields $\widetilde{S}\ $and $%
\ \widetilde{Q}$ \ are introduced:
\begin{eqnarray}
\mathcal{J} &=&\int_{\omega }\sum_{\mathbf{k}}\left\{ \Gamma \tilde{S}_{%
\mathbf{k},\omega }\tilde{S}_{-\mathbf{k},-\omega }+\Theta k^{2}\widetilde{Q}%
_{\mathbf{k},\omega }\widetilde{Q}_{-\mathbf{k},-\omega }-\widetilde{Q}_{%
\mathbf{k},\omega }\left[ (-\omega ^{2}+i\Theta k^{2}\omega )Q_{-\mathbf{k}%
,-\omega }+\frac{\partial \mathcal{H}}{\partial Q_{\mathbf{k},\omega }}%
\right] \right.  \notag \\
&&\left. -\tilde{S}_{\mathbf{k},\omega }\left( i\omega S_{-\mathbf{k}%
,-\omega }+\Gamma \frac{\partial \mathcal{H}}{\partial S_{\mathbf{k},\omega }%
}\right) \right\} .
\end{eqnarray}%
The magnetic linear response function, for example, is defined as the
derivative $\langle S(t)\rangle $ over the external field $h(t^{\prime })$:
\begin{equation}
\frac{\delta {\large {\langle }}S(t){\large {\rangle }}}{\delta h(t^{\prime
})}=\Gamma {\large {\langle }}S(t)\tilde{S}(t^{\prime }){\large {\rangle ,}}
\end{equation}%
where angular brackets denote nonequilibrium average \cite{Janss79,Domin77}
\begin{equation}
{\large {\langle }}O(S,\tilde{S},Q,\widetilde{Q}{\large {)\rangle }}=\frac{1%
}{Z}\int {\large {\mathcal{D[}}}i\tilde{S}{\large {\mathcal{]}}}{\large {%
\mathcal{D[}}}S{\large {\mathcal{]D[}}}i\widetilde{Q}{\large {\mathcal{]}}}%
{\large {\mathcal{D[}}}Q{\large {\mathcal{]}}}\ O[S,\tilde{S},Q,\widetilde{Q}%
]\ \exp \mathcal{J}\{\tilde{S},S,Q,\widetilde{Q}\}.
\end{equation}%
In this formalism we have two kinds of propagators: the free response
propagators and the free two-point correlation functions \cite%
{IroSchwabl83,Deng87,Paw98}.

Next, using dynamic Gaussian transformations we decouple the sound mode from
the order parameter getting an expression for the acoustic self-energy \cite%
{IroSchwabl83,Deng87,Paw98}
\begin{equation}
\Sigma _{ac}(\emph{k},\omega )=2f_{0}^{2}\emph{k}^{2}\Pi (k,\omega ),
\label{sigma3}
\end{equation}%
where the four-spin response function $\Pi (k,\omega )\equiv \langle \Gamma
\widetilde{S^{2}}_{\mathbf{-k},-\omega }S_{\mathbf{k},\omega }^{2}\rangle ^{%
\mathcal{J}_{A}}$ is calculated with the effective, phonon-free dynamic
functional $\mathcal{J}_{A}$ of the model A in the classification of
Halperin and Hohenberg \cite{HH77,Paw98}. Here $\widetilde{S^{2}}_{\mathbf{k}%
}(t)=\frac{1}{\sqrt{V}}\sum_{\mathbf{k}_{1}}\tilde{S}_{\mathbf{k}_{1}}(t)S_{%
\mathbf{k}-\mathbf{k}_{1}}(t)$ plays the role of reaction field coupled to
the square of spin. The sound attenuation coefficient is given by the
imaginary part of the acoustic self-energy
\begin{equation}
\alpha (\omega )=\frac{1}{2c_{0}\omega }\func{Im}\Sigma _{ac}(\frac{\omega }{%
c_{0}},\omega )\simeq \frac{f_{0}^{2}\omega }{c_{0}^{3}}\func{Im}\Pi
(0,\omega ),  \label{alpha}
\end{equation}%
and the sound velocity by the real part
\begin{equation}
\frac{c(\omega )-c_{0}}{c_{0}}\simeq -\frac{f_{0}^{2}}{c_{0}^{2}}\func{Re}%
[\Pi (0,\omega )],  \label{Vel}
\end{equation}%
where $c_{0}$ is the bare sound velocity. In Eqs. (\ref{alpha}) and (\ref%
{Vel}) the ultrasonic wave vector \ $\mathbf{k}$ \ has been put to zero in $%
\Pi (k,\omega )$ as in the ultrasonic experiments the wavelength is much
greater than the \ correlation length \ $\xi $ so we can take $k\xi =0$. In
Eq.(\ref{alpha}) the noncritical term $\Theta \omega ^{2}/2c_{0}^{3}$ has
been omitted. \

In this model we have considered the isotropic solid for simplicity. Although
most of the crystals are anisotropic but, as was shown by Schwabl and Iro \cite%
{IroSchwabl83}, the acoustic self-energy is still given by Eq. (\ref{sigma3}%
)  provided the isotropic coupling constant $f_{0}$ and bare velocity $c_{0},$
are replaced by $f_{0}(\hat{k},\lambda )$ and \ $c_{0}(\hat{k}%
,\lambda ),$ respectively, where $\hat{k}=\vec{k}/k$ is the direction of
propagation and $\lambda $ the polarization of the sound mode. In high \
symmetry directions of propagation the coefficient  $f_{0}(\hat{k},\lambda )$  is different from zero only
for the longitudinal modes \cite{Deng87}.

\section{Critical attenuation}

\begin{figure}[tb]
\begin{center}
\vskip 4pt \includegraphics[
height=1.8cm,width=8cm ]{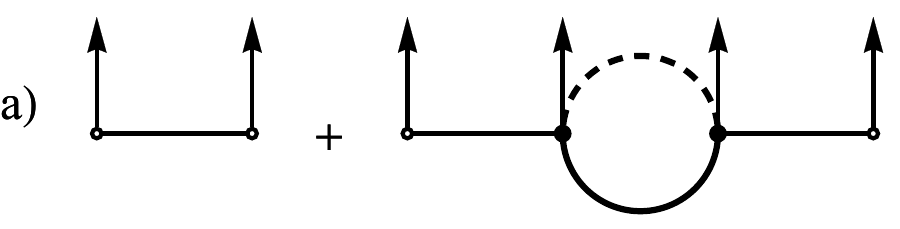}
\includegraphics[
height=1.8cm,width=8cm ]{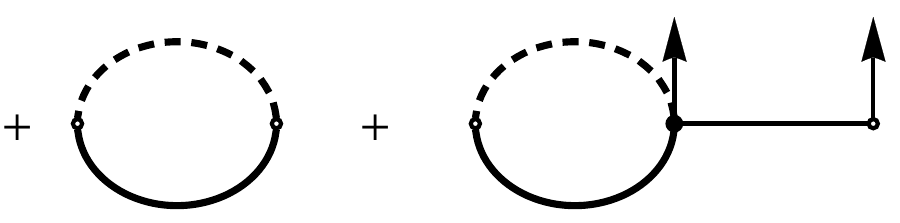}
\end{center}
\par
\vskip 14pt
\begin{minipage}{0.5\columnwidth}
        \centering\includegraphics[width=.97\columnwidth]{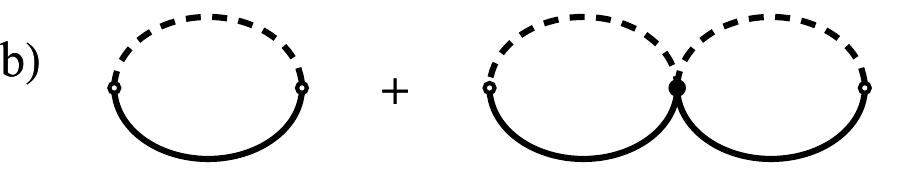}
  \end{minipage}%
\begin{minipage}{0.5\columnwidth}
        \centering\includegraphics[width=.5\columnwidth]{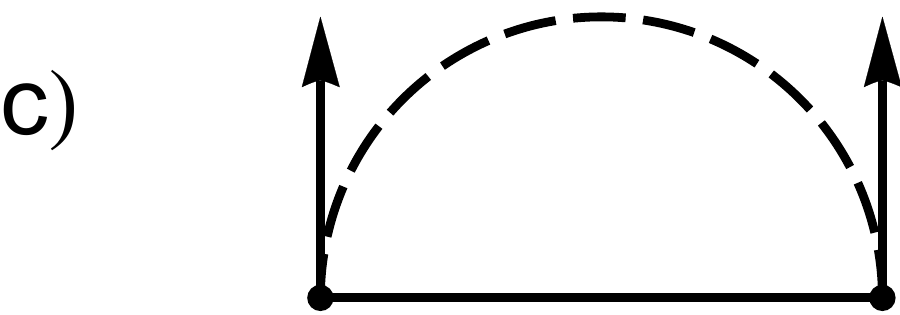}
   \end{minipage} \vspace{1cm}
\caption{a) \ One-loop diagrams for the acoustic self-energy. Full lines
represent the spin response function $\langle \widetilde{S}\protect\delta %
S\rangle $ and the dashed lines represent the correlation function \ $%
\langle \protect\delta S\protect\delta S\rangle $. The arrows represent the
static magnetization $M$ and the black points the coupling constant $u$. b)\
\ The two-loop approximation for $\Sigma _{ac}$ in the disordered phase. c)\
\ One-loop diagram for the self-energy of the spin response function. }
\label{Fig1}
\end{figure}

Putting $S=M+\delta S,$ \ where $M=\langle S\rangle $ is the equilibrium
value of magnetization and $\delta S$ is the spin fluctuation, we can write
the four-spin response function as \cite{Deng87}
\begin{equation}
\Pi =\langle (\Gamma \widetilde{S}\delta S)(\delta S\delta S)\rangle
+2M\langle (\Gamma \widetilde{S}\delta S)\delta S\rangle +M\langle \Gamma
\widetilde{S}(\delta S)^{2}\rangle +2M^{2}\langle \Gamma \widetilde{S}\delta
S\rangle .  \label{Pi1}
\end{equation}%
The last term known also as a relaxational term is the analogue of the
Landau-Khalatnikov sound damping \cite{LK54} which is the only one which
contributes to sound attenuation\ in the mean-field theory. \ The first term
in Eq.(\ref{Pi1}) is known as a fluctuation contribution to the sound
attenuation and the other two terms are sometimes called the mixing
contribution \cite{Foss85}. In the high-temperature phase only the four-spin
response function $\langle (\widetilde{S}\delta S)(\delta S\delta S)\rangle $
has to be obtained. The critical exponent of this 'energy' response function
is equal to the specific-heat exponent $\alpha .$\ \ The theory becomes more
complicated as there is non-zero magnetization. \ Considered separately, the
three contributions are characterized by different critical exponents and
for example the Landau-Khalatnikov term diverges with the critical exponent
equal to $2\gamma -2\beta $ \cite{Foss85}, \ where $\gamma $ and $\beta $
are the susceptibility and order parameter exponents, respectively. However,
as was noted by Halperin and Hohenberg \cite{HH77} \ in the scaling region
there should be cancelations between different contributions and the
critical sound attenuation exponent should be the same as in the disordered
phase. \ It was pointed later explicitly by Dengler and Schwabl \cite{Deng87}
that this was the case. To evaluate $\Pi $\ \ in the critical region several
RG methods are at disposal. It is well known that calculations of dynamical
functions are rather complicated because we need to exponentiate
singularities at both large and small arguments \cite{Nelson76}. Many
alternative approaches can give different results depending on the
exponentiation procedure and the region of the parameters $\omega ,~\ t$ and
$h$. We use here the direct perturbational method based on the $\epsilon
=4-d $ \ expansion \cite{Wilson72,Comb75} with sharp cutoff $\Lambda =1$ and
coupling constants chosen equal to their fixed-point values $u=u^{\ast }=$O$%
(\epsilon )~,v=0$. The one-loop diagrams contributing to \ \ $\Pi $\ \ are
shown \ in Fig. \ref{Fig1}a.
Exponentiation of \ logarithms \ \cite{Wilson72,Comb75}\ gives
\begin{equation}
\Pi =\chi ^{\alpha /\gamma }f(y,s),  \label{Phi2}
\end{equation}%
where $\chi (t,h)$ is the static susceptibility and $\ f$ \ is a scaling
function depending on the \ reduced frequency $y=\omega /(\Gamma b\chi
^{-z\nu /\gamma })$, and on a scaling variable \ $s=2tm^{-1/\beta }$\ ,
where $m^{2}=2u^{\ast }M^{2}$ and \ $z=2+c\eta $ \ is a dynamic exponent for
the universality class A \ (with a constant $c=$O$(1))$ \ \cite{HH77}. \
Here\ \ $\beta ,\gamma ,\nu $ and $\eta =\mathrm{O}(\epsilon ^{2})$ are
usual static critical exponents. For calculating the attenuation scaling
function we used a popular approximation: $f=A_{\text{rel}}f_{\text{rel}}+$\
$A_{\text{fluc}}f_{\text{fluc}}$\ \ where $A_{\text{rel}}$ \ and $A_{\text{%
fluc}}$ are critical amplitudes and the relaxational and fluctuation scaling
function are found to first order in $\epsilon $:%
\begin{equation}
f_{\text{rel}}(y,s)=P(s)^{2\beta /\nu }\frac{1-6u^{\ast }\unit{K}_{d}\pi
_{1}(y)}{1-iy-9u^{\ast }\unit{K}_{d}\pi _{2}(y,s)},  \label{Frel}
\end{equation}%
\begin{equation}
f_{\text{fluc}}(y)=\frac{iz\nu }{2\alpha y}\left[ \left( 1-\frac{iy}{2}%
\right) ^{1-\frac{\alpha }{z\nu }}-1\right] +O(\epsilon ^{2}),  \label{Ffluc}
\end{equation}%
where $\pi _{1}(y)=-\frac{i}{y}(1-\frac{iy}{2})\ln (1-\frac{iy}{2})$ and $%
\pi _{2}(y,s)=-(\frac{2}{s+3})[-\frac{1}{2}+\frac{iy}{8}+\frac{i}{y}(1-\frac{%
iy}{2})\ln (1-\frac{iy}{2})$ $\ $with $\unit{K}_{d}=2^{-d+1}\pi
^{-d/2}/\Gamma (d/2)\ $. \ \ The function  \ $f_{\text{rel}}$ corresponds to the
Landau-Khalatnikov relaxational contribution. The factor $P(s)^{2\beta /\nu }$ \ in Eq. (\ref%
{Frel}) is a consequence of \ $M^{2}$ factor in the last term in Eq. (\ref%
{Pi1}).\ \ The function $\pi _{2}(y,s)$ describes contribution from the
second diagram in Fig. \ref{Fig1}a which is responsible for renormalization
of the bare magnetic response function by the only O($\epsilon $) diagram
contributing to the magnetic self-energy shown in Fig. \ref{Fig1}c.\ \ The
forth diagram in Fig. \ref{Fig1}a \ give an O($\epsilon $) correction to the
Landau Khalatnikov theory which is proportional to the function $\pi _{1}$
in Eq.(\ref{Frel}). In the order O($\epsilon $) \ considered here, there is
no need to exponentiate logarithms like $\ln (1-\frac{iy}{2})$ in \ $f_{%
\text{rel}}(y)$ \ due to the fact that with increasing $y$ the leading
Landau-Khalatnikov term vanishes more rapidly than the first-order
correction \cite{Deng87}. In order to obtain the function $f_{\text{fluc}%
}(y) $ also a two-loop diagram shown in Fig \ref{Fig1}b needs to be
considered. The most important in the exponentiation procedure is that in
order to obtain the correct asymptotic behavior for $y\rightarrow \infty $\
\ \ the exponent $\epsilon /12$ has been replaced by \ $\alpha /2z\nu $ $\ $%
\cite{Deng87}. \

The normalization factor $b(s)=1+\frac{\epsilon }{8}P(s)^{2\beta /\nu }$,\ \
is introduced here for the normalization reason and follows from the low $%
\omega $ expansion of the dynamic order-parameter susceptibility \cite%
{Calab,PawErd11,PawErd13}. The function \ $P(s)=m^{2}\chi ^{2\beta /\gamma
}\simeq (\frac{2}{s+3})^{\nu }[\frac{s}{s+3}+\frac{3}{s+3}(\frac{2}{s+3})^{%
\frac{\nu }{2\beta }-\nu }]$ \ \ may be thought of as\ \ a crossover profile
which is equal to $0$ for $s\rightarrow \infty $ (in the paramagnetic phase)
and $1$ in the coexistence region ($s=-1$) and changes to \ $\left( \frac{2}{%
3}\right) ^{\nu /2\beta }$\ for the critical regime, $s\rightarrow 0$. \ \
Note that this O($\epsilon $) correction vanishes for the disordered phase $%
s\rightarrow \infty .$ \

\section{Landau-Ginzburg theory}

\begin{figure}[t]
\begin{center}
\begin{minipage}{0.8\columnwidth}
        \centering\includegraphics[width=.75\columnwidth]{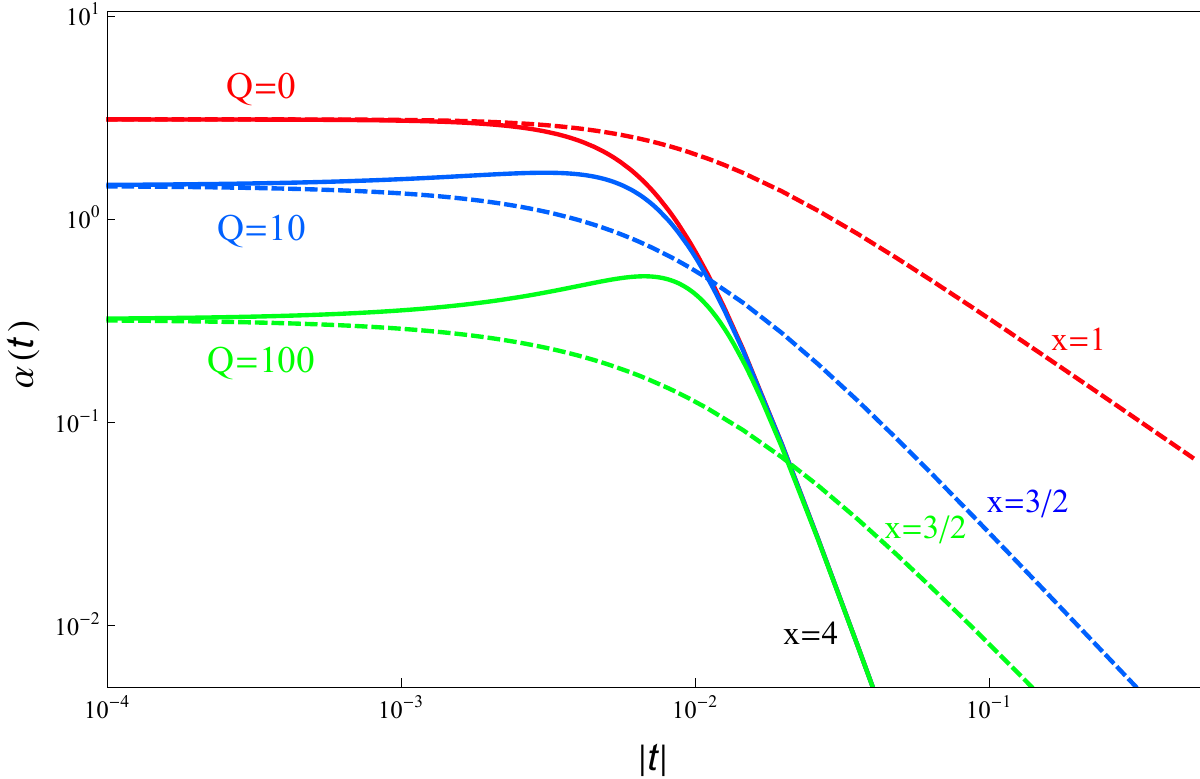}
  \end{minipage}
\end{center}
\par
\vspace{-0.5cm}
\caption{The typical ultrasonic attenuation coefficient temperature
dependence in the Landau-Ginzburg theory for three values of $Q$. Non-zero
magnetic field $h=1$ Oe is assumed. The continuous lines correspond to $%
T>T_{C}$ and the dashed lines to $T<T_{C}$. The hydrodynamic sound
attenuation exponent $x$ is also shown.}
\label{Fig2}
\end{figure}

If the fluctuations of the order-parameter are neglected only the
relaxational term remains:%
\begin{equation}
\Sigma _{ac}(k,\omega )=4f_{0}^{2}k^{2}M^{2}\chi (\omega
,t)=4f_{0}^{2}k^{2}M^{2}\frac{\chi _{GL}}{1-iy_{GL}},  \label{SigMFA}
\end{equation}%
where $y_{GL}=\omega \tau _{GL}$ is the reduced frequency \ with $\tau
_{GL}=\chi _{GL}/\Gamma $\ $\ $as a characteristic relaxation time and$\
\chi _{GL}$ as the static susceptibility in the Landau-Ginzburg theory. In
order to calculate the equation of state as well as the susceptibility, $%
\chi _{GL}=(\partial ^{2}\Phi _{\mathrm{GL}}/\partial M^{2})^{-1},$\ \ the
Landau-Ginzburg functional is postulated in the form of regular expansion%
\begin{equation}
\Phi _{\mathrm{GL}}(M)=\frac{1}{2}r_{0}M^{2}+\frac{u}{4}M^{4}+\frac{v}{6}%
M^{6}+~...~-Mh  \label{GL}
\end{equation}%
In many phenomenological approaches the sixth and higher-order terms in Eq. (%
\ref{GL}) are omitted, \ however as was shown by Kuz'min \cite{Kuz08} the
sixth-order term is essential for finding an appropriate equation of state
for ferromagnets, which is valid for an arbitrary magnetic field $h$ and
reduced temperature $t$. At thermal equilibrium \ $\Phi _{\mathrm{GL}}$ must
be minimum%
\begin{equation*}
\frac{\partial \Phi _{\mathrm{GL}}(M,h,t)}{\partial M}=0,
\end{equation*}%
so the equation of state is written as
\begin{equation}
h=M(at+uM^{2}+vM^{4}),  \label{ESMF}
\end{equation}%
and the susceptibility \ as $\chi _{GL}=(at+3uM^{2}+5vM^{4})^{-1}$. A
log-log plot of \ Eq. (\ref{ESMF}) for $h=0$ shows that \ the effective
exponent $\beta ~$crosses over from typical Landau theory value $1/2$ to $%
1/4~$as the temperature decreases from $T_{C}$ \cite{Kuz08}.\ \ The
proximity to the tricritical point ($u=0$) explains $\beta \approx 1/3$
observed in many ferromagnets \cite{Kuz08}.\ \ Recently we have shown \cite%
{Paw20} \ \ that the sixth-order term is also crucial for understanding of
the ultrasonic attenuation in manganese phosphide MnP. It was demonstrated
that the characteristic quotient $Q=v/u$ \cite{Kuz08}$,$ which determines
how far below the Curie temperature the Landau asymptotic behavior for
magnetization $M\sim |t|^{1/2}$ \ still holds ($M\sim |t|^{1/4}$ \ relation
is observed for the tricritical regime), has also dramatic effect on
dynamical properties\ \cite{Paw20} \ and large values of $Q$ \ \ favor the
shift of the maximum of the ultrasonic attenuation to higher than $T_{C}$
temperatures as is seen also in the ultrasonic data in MnP \cite%
{Kom74,Tach74}. The imaginary part of $\ \Sigma _{ac}(k,\omega )$\ \ is the
source of ultrasonic attenuation and from Eq. (\ref{SigMFA}) \ we obtain
\begin{equation}
\alpha (\omega ,t,h)=\frac{2f_{0}^{2}\omega ^{2}}{c_{0}^{3}\Gamma }\frac{%
M^{2}(t,h)\chi _{GL}(t,h)^{2}}{1+y_{GL}^{2}(\omega ,t,h)}.  \label{alfaMFA}
\end{equation}%
The product $M^{2}\chi _{GL}^{2}$ determines the sound attenuation exponent:
$\alpha \sim |t|^{-x_{\pm }}$ in the hydrodynamic regime, where $\omega \tau
_{GL}\ll 1$. The index $+$ refers to $T>T_{C}$ \ end \ $-$ \ to $T<T_{C}$. \
This exponent determines the slope of the attenuation peak in the log-log
plot. \ It is clear by a simple inspection of \ Eq. (\ref{alfaMFA}) that $%
x_{-}=1$ for the ordinary Landau theory behavior \ ( $Q=0$) and $x_{-}=3/2$
for the tricritical behavior ( $Q=$ $\infty $). \ So, in the Landau-Ginzburg
theory the effective sound attenuation exponent\ \ $x_{-}^{eff}$\ \cite%
{Riedel74,Paw09} \ \ in the low-temperature phase may take an intermediate
value between $1$ and $3/2$ depending on the value of \ $Q$ and the range of
the reduced temperature. It is obvious that \ \ $x_{+}=0$ \ for $h=0$ in
this theory as $M=0~$ for \ \ $T>T_{C}$. For non-zero magnetic field however
$M\neq 0$ even for $T>T_{C}$ \ and \ \ \ much larger attenuation exponent $%
x_{+}=4$ is obtained. \ It is consequence of the relation $M=\chi _{GL}h$ \
for small magnetization, where the Landau-Ginzburg susceptibility obeys $%
|t|^{-1}$ \ power-law in both: Landau theory and the tricritical limits. So
for finite magnetic field the sound attenuation coefficient \ curve has much
steeper slope for $T>T_{C}$ than in the low-temperature phase as was shown
in Fig. \ref{Fig2}. From Fig. \ref{Fig2} \ one can also observe that the
displacement of the sound attenuation peak towards the higher temperatures\
is monotonic with respect to the value of the characteristic quotient $Q.$ \

A crossover from mean-field to tricritical behavior explains the ultrasonic
data in MnP for moderate and high magnetic fields \cite{Paw20}\ but does not
explains the data in the region near the critical point called also the
fluctuation region. For example, $\alpha (T>T_{C},h=0)$ is equal zero in
Landau-Ginzburg theory contrarily to what is observed in experiment. The
fluctuations has to be taken into account in MnP at least in the region of
low magnetic fields and temperatures very close to the critical point.

\section{Crossover parametric model}

As regards the explicit representation of the critical behavior it is
frequently more convenient to express everything in the parametric
representation of an asymptotic equation of state. The reason for that is
that many scaling functions are given by various approximations of
renormalization group theory\ and therefore are sometimes very cumbersome
and inconvenient for practical use \cite{Aga01}. \ Usually the starting
point is a simple linear model originally introduced by Schofield \cite%
{SchLitHo69,Sch69,Jos69}. It is defined by:
\begin{equation}
h=h_{0}r^{\beta \delta }\theta (1-\theta ^{2})\equiv h_{0}r^{\beta \delta
}l(\theta ),  \label{hh}
\end{equation}%
\begin{equation}
t=r(1-b^{2}\theta ^{2})\equiv rk(\theta ),  \label{tt}
\end{equation}%
\begin{equation}
M=M_{0}r^{\beta }\theta  \label{mm}
\end{equation}%
where $h_{0}$, $M_{0}$ and $b^{2}$ are constants \ and $r$ and $\theta $ are
abstract parametric variables chosen such that $\theta =\pm 1$ correspond to
the two branches of the coexistence curve, $\theta =1/b$ to the critical
isotherm and $\theta =0$ to the critical isochore. The parameter $r$ \ in
non-negative and measures a "radial" distance from the critical point $\ $%
and $\ \theta $ \ (pseudoangle) specifies the location on the contour of
constant $r$.\ In this representation all critical singularities are
incorporated as power laws in the variable $\ r$, while the dependence on $\
\theta $\ \ is kept analytic.\ \ The great advantage of the linear
parametric representation is that it generates a closed form expressions for
all thermodynamic functions.\ \ The resulting parametric expressions for a
number of thermodynamic properties are given in Appendix A. For example the
susceptibility is given by%
\begin{equation}
\chi =\frac{M_{0}}{h_{0}}r^{-\gamma }\frac{k-\beta k^{\prime }\theta }{%
kl^{\prime }-\beta \delta k^{\prime }l}.
\end{equation}%
The quantities with a prime indicate the derivative with respect to $\theta $%
. The constants $h_{0}$ and $M_{0}$\ are determined by fitting the
experimental data. \ According to the universality of the critical phenomena
the constant $b$\ \ \ is assumed to be universal and may be \ fixed by
matching the ratios of critical amplitudes obtained within the parametric
representation to the known ratios obtained, for example, from the
renormalization group theory or \ Monte-Carlo simulations. A satisfactory
fit to the amplitude ratios is provided by taking $b^{2}=1.30$ \ \cite%
{Borjan} and this value is adopted within present study.\ The linear model
appears to be consistent with the asymptotic renormalization group theory to
second order in $\epsilon $ \cite{WallZia74}.

Various improvements of the linear model has been proposed \cite%
{Aga01,FishZinn}. \ \ Many of them concentrate on the critical region with
inclusion of the Wegner correction--to-scaling \ contributions \cite{Weg72}
and the others on the crossover from the asymptotic critical region to the
regular classical (mean-field) behavior far away from the critical point
\cite{Luett92}. \ \ In the approach presented by Agayan et al. \cite{Aga01}
a crossover parametric transformation similar to the one deduced for the
renormalization-group theory \cite{Nic81,Nic85,Nic86,Chen1,Chen2} \ was
proposed. According to non-asymptotic renormalization group procedure and
implementing a matching point proposed by Nicol and co-workers \cite%
{Nic81,Nic85,Nic86} Chen et al. \cite{Chen1,Chen2} put forward the following
crossover expression for the singular part of the actual\ Helmholtz \ free
energy in the Ising system:%
\begin{equation}
\Delta A_{s}=\frac{1}{2}tM^{2}\mathcal{TD+}\frac{\overline{u}u^{\ast
}\Lambda }{4!}M^{4}\mathcal{D}^{2}\mathcal{U}-\frac{1}{2}t^{2}\mathcal{K},
\label{Across}
\end{equation}%
where \ $\overline{u}=u/u^{\ast }$\ is a rescaled coupling constant \ and $%
\mathcal{T}$,$\mathcal{D}$,$\mathcal{U}$ and $\mathcal{K}$ are rescaling
functions defined by%
\begin{equation}
\mathcal{T}=Y^{(2\nu -1)/\Delta _{s}},~\mathcal{D}=Y^{-\eta \nu /\Delta
_{s}},\mathcal{U}=Y^{\nu /\Delta _{s}},
\end{equation}%
and%
\begin{equation}
\mathcal{K}=\frac{\nu }{\alpha \overline{u}\Lambda }(Y^{-\alpha /\Delta
_{s}}-1).
\end{equation}%
Note that in Eq. (\ref{Across}) the substitution $at\rightarrow t,$ $v=0$
and $\frac{u}{4}\rightarrow \frac{\overline{u}u^{\ast }\Lambda }{4!}$ was
performed compared to our starting Landau-Ginzburg Hamiltonian $\mathcal{H}_{%
\mathrm{OP}}$. The parameter \ $\Lambda $ \ introduced here for later
convenience is to be interpreted as a dimensionless wave number related to
the actual cutoff wave number \cite{Nic81,Aga01}. The crossover function
\cite{Chen1,Chen2} $Y\equiv \frac{u(l)-u^{\ast }}{u-u^{\ast }},$where \ $l$
\ is a rescaling parameter, \ calculated at the matching-point $l=l^{\ast }$
\begin{equation}
\Lambda e^{-l^{\ast }}=\kappa ,
\end{equation}%
is given by the relation \cite{Chen1,Chen2,Aga01}:
\begin{equation}
1-(1-\overline{u})Y=\overline{u}(1+\frac{\Lambda ^{2}}{\kappa ^{2}}%
)^{1/2}Y^{\nu /\Delta _{s}},  \label{Y}
\end{equation}%
with $\kappa $ (a parameter proportional to the inverse of correlation
length, which is a measure of the distance to the critical point) \ defined
by
\begin{equation}
\kappa ^{2}=t\mathcal{T}+\frac{1}{2}\overline{u}u^{\ast }\Lambda M^{2}%
\mathcal{DU}.  \label{kapp}
\end{equation}%
\ \ When $\kappa $ is small the crossover function $Y$ approaches zero like
\ $(\kappa /\overline{u}\Lambda )^{\Delta _{s}/\nu }$ and the asymptotic
power-laws are recovered from Eq. (\ref{Across}) including the leading
Wegner corrections to scaling. The classical limit corresponds to $\overline{%
u}\Lambda /\kappa \rightarrow 0$ or $Y\rightarrow 1$ \ and the
Landau-Ginzburg expansion is easily recovered \ from Eq.(\ref{Across}) as
the rescaling functions $\mathcal{T},\mathcal{D},\mathcal{U}$ \ tend to
unity while $\mathcal{K}\rightarrow 0$. \ In Eq.(\ref{Across}) higher-order
terms can be also included, for example the contribution from the $\frac{v}{6%
}M^{6}\mathcal{\ }$\ in Eq.(\ref{GL})\ \ term transforms to $\frac{v}{6}M^{6}%
\mathcal{D}^{3}\mathcal{U}^{3/2}$ in the crossover Helmholtz \ free energy
\cite{Chen2,Luett92}. \

The asymptotic parametric models (including described here the simplest
linear model) are valid only in the vicinity of the critical point. \ In
order to built a crossover parametric model \ able to describe the crossover
from critical to mean-field behavior Agayan et al. \cite{Aga01} \ related
the distance parameter $r$ \ in parametric representation \ to the inverse
of correlation length by equation%
\begin{equation}
\kappa ^{2}(r)=arY^{(2\nu -1)/\Delta _{s}},  \label{par1}
\end{equation}%
which reflects the observation that the parameter $\kappa ^{2}$\ \ \ plays a
similar role to the distance variable $r$ ($\kappa \rightarrow 0$ near the
critical point and $\kappa \rightarrow \infty $ far away from the critical
point). It is easy to see from Eq. (\ref{par1})\ \ that $\kappa \sim r^{\nu
} $ near the critical point and \ $\kappa \sim r^{1/2}$\ far away from the
critical point. The crossover function is again given in the implicit form
by Eq. (\ref{Y}) \ and can be rewritten as%
\begin{equation}
1-(1-\overline{u})Y(r)=\overline{u}(1+\frac{\Lambda ^{2}}{arY(r)^{(2\nu
-1)/\Delta _{s}}})^{1/2}Y(r)^{\nu /\Delta _{s}}.  \label{Ybis}
\end{equation}%
The crossover function $Y$ is only a function of $r$ and is independent from
the angle variable $\theta .$ \ It also depends on two parameters $\overline{%
u}$ and $\Lambda ^{2}/a$\ \ called also crossover variables \cite{Aga01} \
which \ \ may be related to the Ginzburg number $N_{G}$:%
\begin{equation}
N_{G}=n_{0}g=n_{0}\frac{(\overline{u}\Lambda )^{2}}{a},  \label{Gi}
\end{equation}%
where $n_{0}$ is a constant.\ In the asymptotic critical limit $Y\rightarrow
(r/g)^{\Delta _{s}}$ \ where $g=(\overline{u}\Lambda )^{2}/a$ is again a
crossover parameter proportional to the Ginzburg number.\ \ \ Far away from
criticality \ $Y\rightarrow 1$.

To completely define the crossover parametric model one need to specify the
equations for $h,$ $t$ \ and $M$. For this purpose we modify the linear
model equations (\ref{hh}) as%
\begin{equation}
h=h_{0}g^{\beta \delta -3/2}r^{3/2}\theta \lbrack Y^{(2\beta \delta
-3)/2\Delta _{s}}(1-\theta ^{2})+vr\theta ^{4}],  \label{hhmod}
\end{equation}%
\begin{equation}
M=g^{\beta -1/2}M_{0}r^{1/2}Y^{(\beta -1/2)/\Delta _{s}}\theta  \label{mmod}
\end{equation}%
while the dependence of \ $t$ on $r$ and $\theta $\ \ \ is left unchanged
(Eq. (\ref{tt})). Note that near the critical \ point \ ($r\rightarrow 0$, $%
Y\rightarrow (r/g)^{\Delta _{s}})$ \ \ the term $vr\theta ^{4}$ in the
square brackets in Eq. (\ref{hhmod}) is very small and can be neglected so
the linear model equations are recovered. The virtue of this parametrization
is that in the other (classical) limit, $r\rightarrow \infty $, \ these
equations reduce to
\begin{equation}
h=\tilde{h}_{0}r^{3/2}\theta \lbrack (1-\theta ^{2})+vr\theta ^{4}],
\label{hclass}
\end{equation}%
\begin{equation}
M=\tilde{M}_{0}r^{1/2}\theta
\end{equation}%
which in turn leads to the very simple Ginzburg-Landau expression for the
equation of state:%
\begin{equation}
h=M[at+u^{\prime }M^{2}+v^{\prime }M^{4}],
\end{equation}%
with $\ \ a=\frac{\tilde{h}_{0}}{\tilde{M}_{0}},$ $\tilde{h}_{0}=g^{\beta
\delta -3/2}h_{0}$, $\ \tilde{M}_{0}=g^{\beta -1/2}M_{0}$, $\tilde{v}%
=vg^{3/2-\beta \delta }$, $\ u^{\prime }=\frac{(b^{2}-1)\tilde{h}_{0}}{%
\tilde{M}_{0}^{3}}$ and $\ v^{\prime }=\frac{v\tilde{h}_{0}}{\tilde{M}%
_{0}^{5}}$. Similar classical expressions are obtained for the
susceptibility and other thermodynamic quantities (see Appendix B).

Recalling the definitions (\ref{sigma3}-\ref{Ffluc}) crossover parametric
attenuation coefficient and dispersion are given by%
\begin{equation}
\alpha (\omega ,r,\theta )=A\omega Y(r)^{-\alpha /\Delta _{s}}\func{Im}\left[
f_{\text{rel}}(y,\theta ,r)+\frac{A_{\text{fluc}}}{A_{\text{rel}}}f_{\text{%
fluc}}(y)\right] ,  \label{alpar}
\end{equation}%
\begin{equation}
c^{2}(\omega )-c^{2}(0)=2Ac_{0}^{3}Y(r)^{-\alpha /\Delta _{s}}\func{Re}\left[
f_{\text{rel}}(y,\theta ,r)-f_{\text{rel}}(0,\theta ,r)+\frac{A_{\text{fluc}}%
}{A_{\text{rel}}}(f_{\text{fluc}}(y)-f_{\text{fluc}}(0))\right]
\label{dyspar}
\end{equation}%
with%
\begin{equation}
f_{\text{rel}}(y,\theta ,r)=\tilde{P}(\theta ,r)\frac{1-6u^{\ast }p(r)\unit{K%
}_{d}\pi _{1}(y)}{1-iy-9u^{\ast }p(r)\unit{K}_{d}\pi _{2}(y,\theta )},
\label{Frelpar}
\end{equation}%
where the function \ $p=u(l)/u^{\ast }$\ \ is given in parametric
representation by
\begin{equation}
p(r)=1-(1-\overline{u})Y(r),  \label{p}
\end{equation}%
and \ \ \ $\pi _{2}(y,\theta )=\theta ^{2}X(\theta ,r)[-\frac{1}{2}+\frac{iy%
}{8}+\frac{i}{y}(1-\frac{iy}{2})\ln (1-\frac{iy}{2})]$. \ \ \ As we approach
the critical point, $r\rightarrow 0,$ \ the parameter $p\ \ $goes to unity $%
\ $and the renormalized coupling constant $u(l)$ approaches the fixed point
value $u^{\ast }.$ \ On the other hand for $r\rightarrow \infty $ it takes
the initial value $u$. \ In Eq. (\ref{Frelpar}) the function $\widetilde{P}%
(\theta ,r)$ corresponds to the factor $P(s)^{2\beta /\nu }$ in Eq. (\ref%
{Frel}). It \ plays important role as it includes all dependence on $\theta $%
\textbf{\ }in the static case ($y=0$). Remembering also that in the statics
the four-spin response function $\Pi $ is proportional to the specific heat\
we took in this paper
\begin{equation}
\tilde{P}(\theta ,r)=\tilde{C}(\theta ,r)-\tilde{C}(0,r)  \label{PP}
\end{equation}%
where $\tilde{C}(\theta ,r)$ is an auxiliary angular function of the
specific heat \ \cite{Aga01}\ which is given in the parametric crossover
model by
\begin{equation}
C_{h}=Y^{-\alpha /\Delta _{s}}\tilde{C}(\theta ,r),  \label{C}
\end{equation}%
with%
\begin{equation}
\tilde{C}(\theta ,r)=\frac{[(1-\alpha Y_{1})\tilde{s}+r\frac{\partial \tilde{%
s}}{\partial r}|_{\theta }](l^{\prime }+5vrY^{\frac{2\beta \delta -3}{%
2\Delta _{s}}}\theta ^{4})-[(\frac{3}{2}+(\beta \delta -\frac{3}{2})Y_{1})l+%
\frac{5}{2}vrY^{\frac{2\beta \delta -3}{2\Delta _{s}}}\theta ^{5}]\tilde{s}%
^{\prime }}{k(l^{\prime }+5vrY^{\frac{2\beta \delta -3}{2\Delta _{s}}}\theta
^{4})-k^{\prime }[(3/2+(\beta \delta -\frac{3}{2})Y_{1})l+\frac{5}{2}vrY^{%
\frac{2\beta \delta -3}{2\Delta _{s}}}\theta ^{5}]},  \label{CC}
\end{equation}%
where \ $\tilde{s}$ \ is an angular function of the entropy (see Appendix B)
and $\frac{\partial \tilde{s}}{\partial r}|_{\theta }$ denotes the radial
derivative of $\tilde{s}$ whereas the primes denote the derivatives with
respect to \ $\theta $ (at constant $r$). In this equation $Y_{1}$ is a
logarithmic derivative of the crossover function%
\begin{equation}
Y_{1}(r)\equiv \frac{d\ln Y}{d\ln r^{\Delta _{s}}}=\frac{1}{\Delta _{s}}%
\frac{r}{Y}\frac{dY}{dr},  \label{Y1}
\end{equation}%
introduced by Agayan et al. \cite{Aga01}. This function approaches unity in
the asymptotic critical limit $r\rightarrow 0$ whereas \ $Y_{1}\rightarrow 0$
far away from the critical point.

The reasons we have chosen the "shape" function $\widetilde{P}(\theta ,r)$
in a form of specific heat are as follows. First, in the limit of low
frequency the response function $\Pi $ \ transforms into the specific heat.
The four-spin response function is sometimes called the frequency-dependent
specific heat. \ It is well known that it plays the essential role in
calculations of sound attenuation and velocity shift in magnets \cite%
{HH77,Kaw68,Paw98,Paw11}. Secondly, Eq. (\ref{PP}) correctly describes also
the mean-field limit \ \ as \ for \ $Y\rightarrow 1,$\ $Y_{1}\rightarrow 0$\
\ we \ see that \ $\tilde{C}(\theta ,r)-\tilde{C}(0,r)\rightarrow \frac{%
\theta ^{2}}{(1-b^{2}\theta ^{2})+3(b^{2}-1)\theta ^{2}+5vr\theta ^{4}}.$
Multiplying \ both the numerator and denominator by $r$ the equation \ (\ref%
{alfaMFA}) \ is recovered.\ \ \ \

It should be noted that also the reduced frequency is changed to\ \ \
\begin{equation}
y=\omega \tau _{0}b(\theta ,r)r^{-1}Y(r)^{\frac{1-z\nu }{\Delta _{s}}%
}X(\theta ,r),  \label{ypar}
\end{equation}%
where $\tau _{0}$ is a bare relaxation time setting the time scale and $%
X(\theta ,r)$ is an angular function of susceptibility \ (Eq. (\ref{B14})).
The frequency normalization factor\ $b(\theta ,r)$\ \ \cite{PawErd13,Calab}\
\ replaces $b(s)$\ \ from Sec. II:\ \
\begin{equation}
b(\theta ,r)=1+\frac{\epsilon }{8}p(r)\theta ^{2}X(\theta ,r).  \label{bpar}
\end{equation}

\section{ Application to \ manganese phosphide $\text{MnP}$}

As an illustration, we show here how the crossover parametric model can be
applied to ultrasonic attenuation data obtained by Komatsubara et al. \cite%
{Kom74} for MnP. We have fitted the experimental data to Eq. (\ref{alpar})
using the critical temperature $T_{C}\ $and the attenuation ($A$), time ($%
\tau _{0}$), temperature ($a$) and magnetic field \ ($h_{0}$) scales\ $\ $as
well as \ $\frac{A_{\text{fluc}}}{A_{\text{rel}}}$\ , $v$, $\overline{u}$ \
and $\Lambda $ \ as adjustable parameters. The values obtained for MnP \ \
are presented in Table I. \ \
\begin{table}[tbh]
\caption{Parameters for MnP from the experimental ultrasonic data
\protect\cite{Kom74}.}
\label{tab1}
\begin{center}
\begin{tabular}{|l|l|l|l|l|l|l|l|l|}
\hline
$\ \ T_{C}$ \ [K] & $\ \ A~\ $[dBcm$^{-1}$] & $\ \ \ \ \tau _{0}$ $\ $[s] & $%
\ \ \ a$ & $\ h_{0}$ [MOe] & $\ \frac{A_{\text{fluc}}}{A_{\text{rel}}}$ & $\
\ \ \ v$ & $\ \ \ \ \overline{u}$ & $\ \ \ \ \Lambda ^{2}$ \\ \hline
$\ ~290.76$ \  &  $\ \ \ 0.0994\ \ $ & $9.08\cdot 10^{-12}$ & $~2.7\ \ $ & $%
\ \ \ \ 0.70\ $ & 0.183 \  & $\ 0.75\ \ $ & $\ 0.089\ \ $ & $\ 0.078\ \ $ \\
\hline
\end{tabular}%
\end{center}
\end{table}
The resulting value of the crossover parameter is%
\begin{equation}
g=6.25\cdot 10^{-4}.  \label{gg}
\end{equation}%
We have adopted the critical exponents for the tree-dimensional Ising-like
systems from the work of \ Pelisseto and Vicari \cite{Peliss}:%
\begin{equation}
\alpha =0.110,\ \ \beta =0.326,\ \ \gamma =1.237,\ \ \delta =4.789,\ \
\Delta _{s}=0.53.
\end{equation}%
\begin{figure}[tbh]
\begin{minipage}{0.5\columnwidth}
        \centering\includegraphics[width=.8\columnwidth]{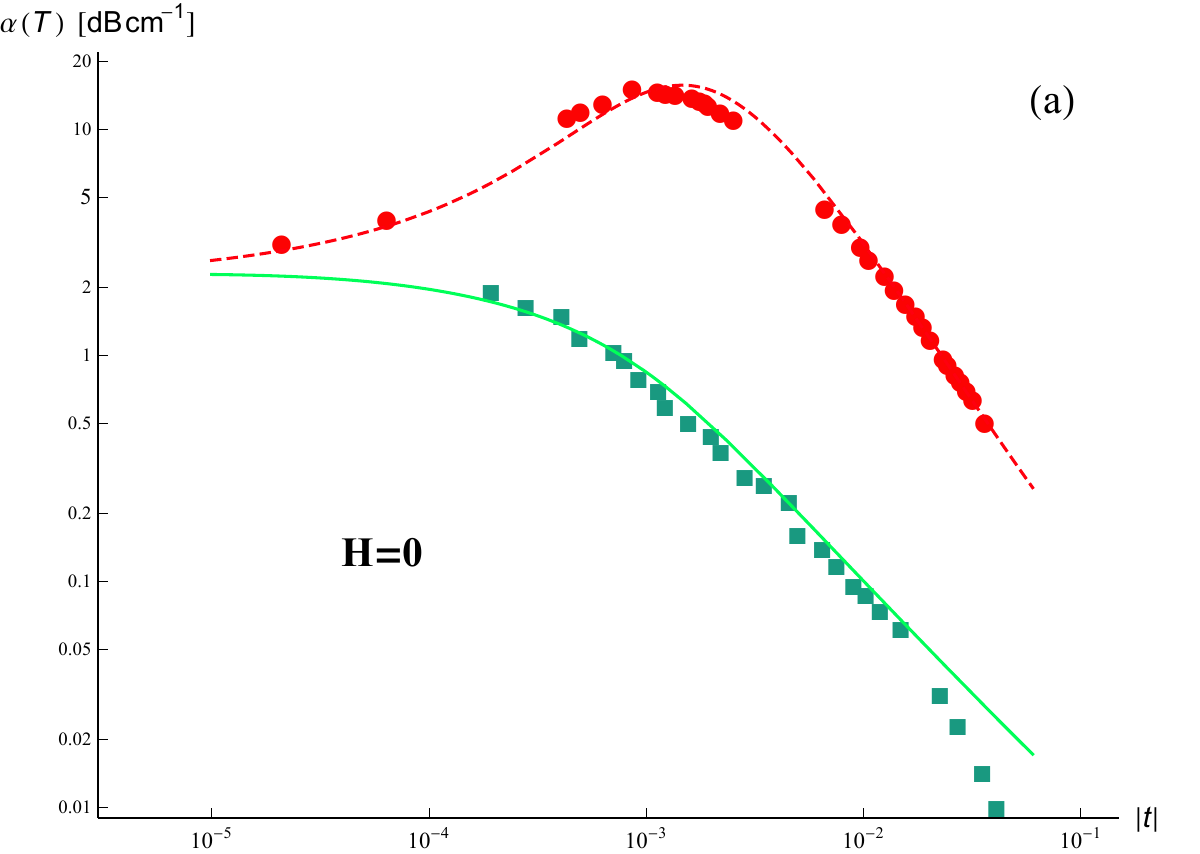}
  \end{minipage}%
\begin{minipage}{0.5\columnwidth}
        \centering\includegraphics[width=.8\columnwidth]{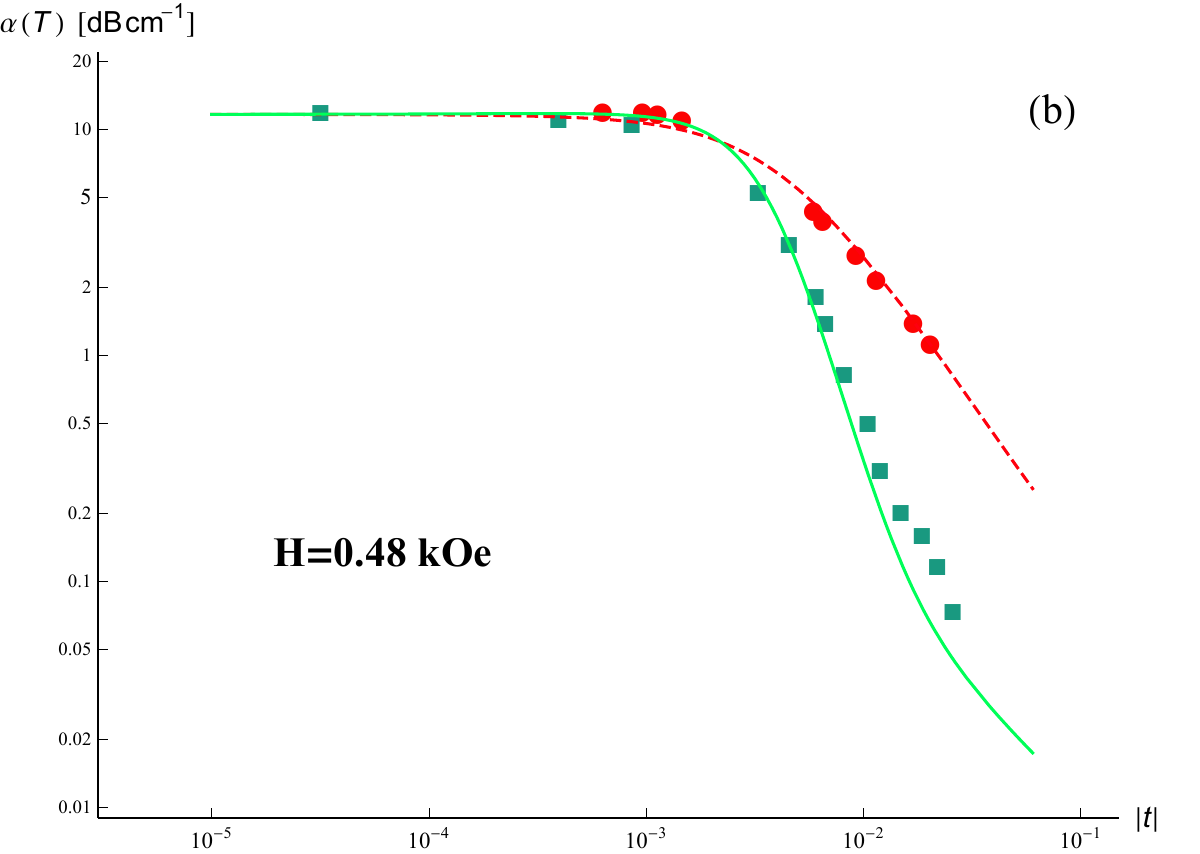}
   \end{minipage} \vspace{1cm}
\begin{minipage}{0.5\columnwidth}
        \centering\includegraphics[width=.8\columnwidth]{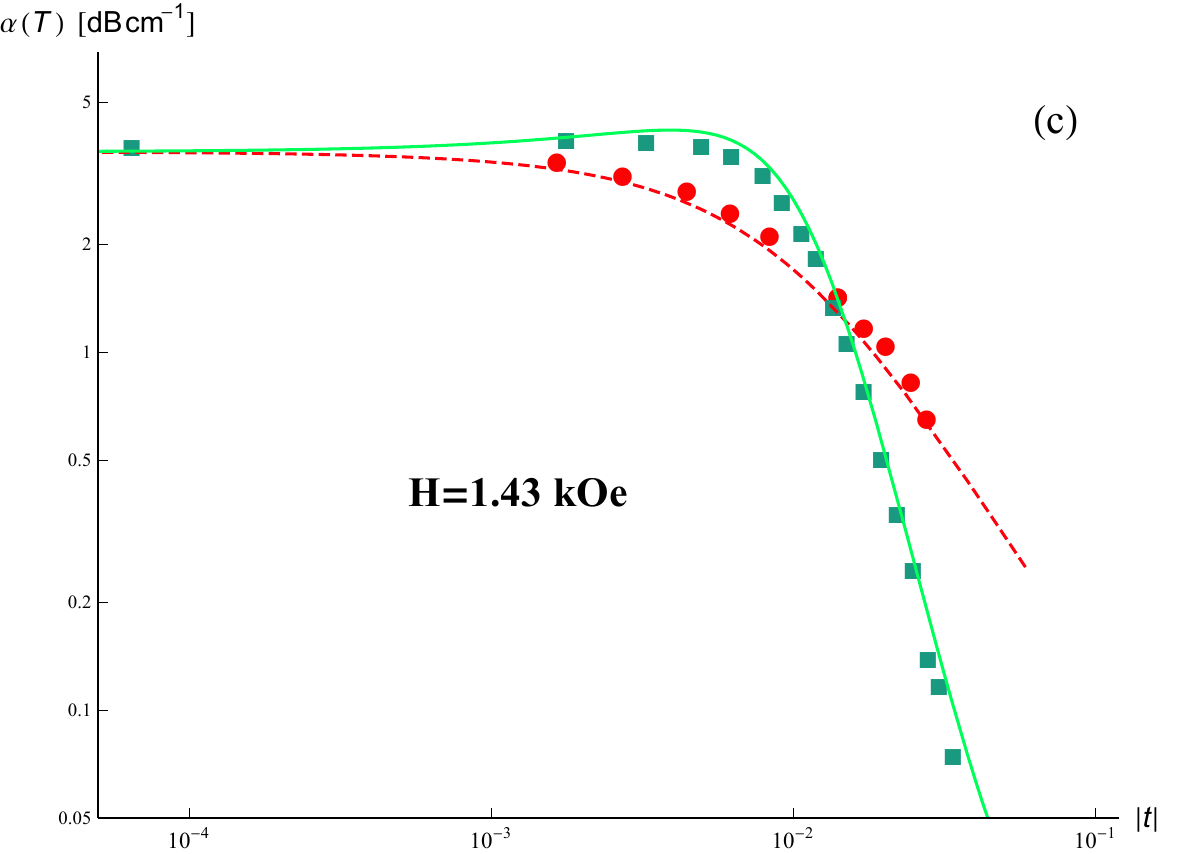}
   \end{minipage}%
\begin{minipage}{0.5\columnwidth}
        \centering\includegraphics[width=.8\columnwidth]{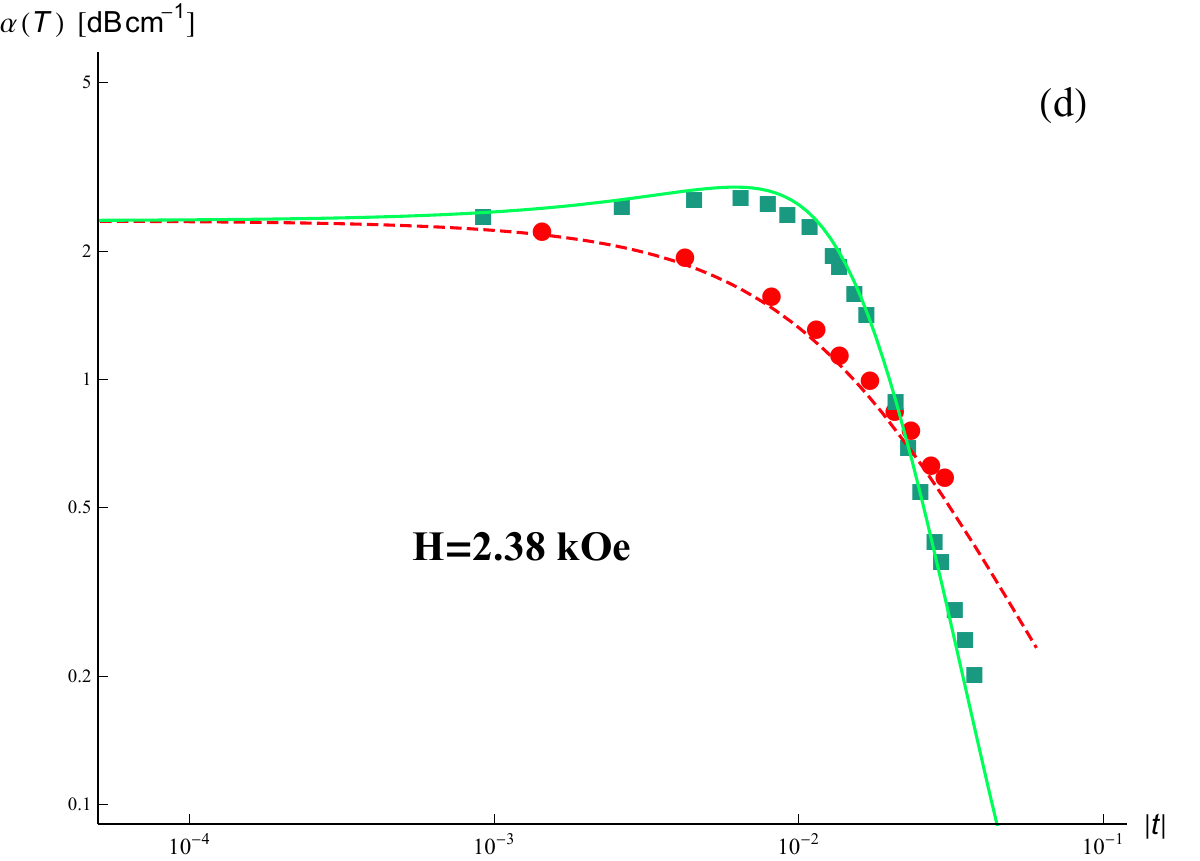}
           \end{minipage}
\begin{minipage}{0.5\columnwidth}
        \centering\includegraphics[width=.8\columnwidth]{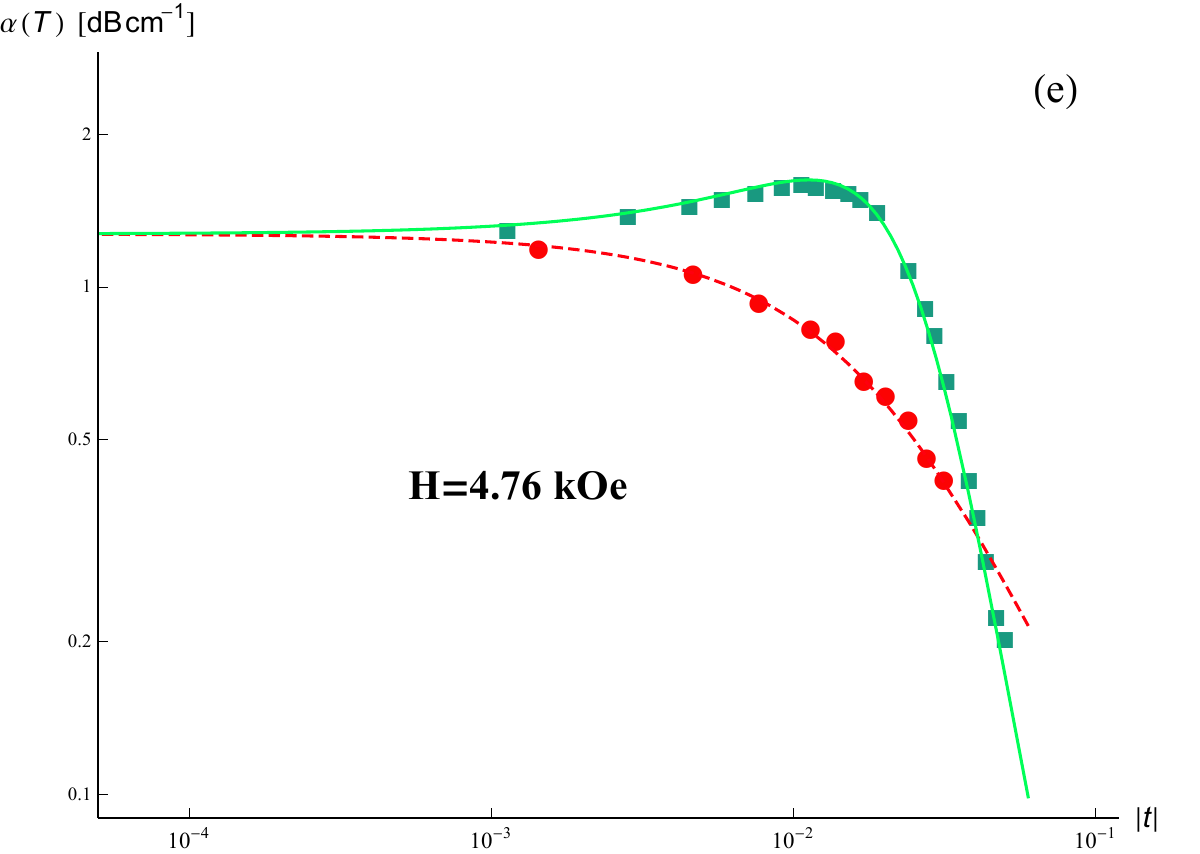}
   \end{minipage}%
\begin{minipage}{0.5\columnwidth}
        \centering\includegraphics[width=.8\columnwidth]{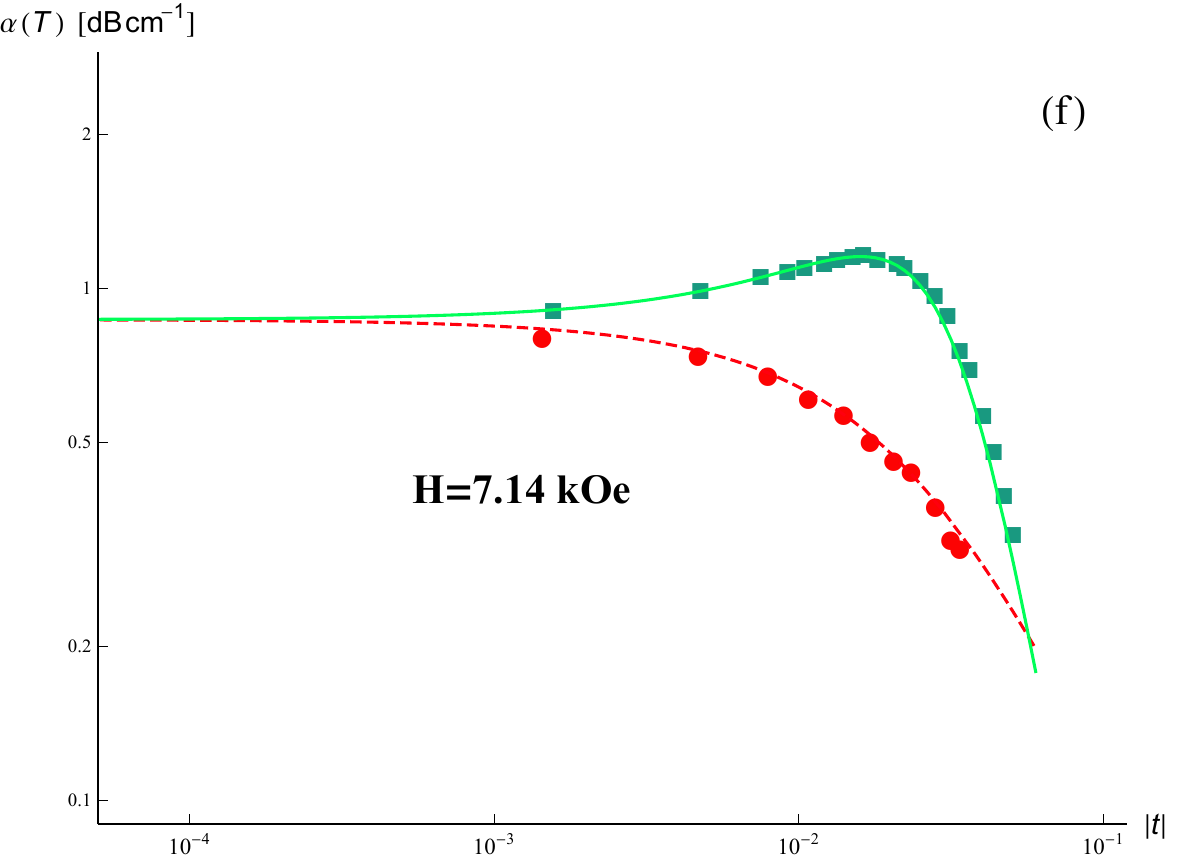}
           \end{minipage}
\caption{Ultrasonic attenuation in MnP. The dashed lines correspond to $%
T<T_{C}$ \ and the continuous lines to $T>T_{C}$. The data are taken from
Ref. \protect\cite{Kom74}.}
\label{Fig3}
\end{figure}
A comparison of \ Eq. (\ref{alpar}) \ with experimental sound attenuation
data in MnP\ is shown in Fig. \ref{Fig3} using a log-log scale. \ Six
different values of magnetic field were shown.The ultrasonic data points are
experimental results obtained by Komatsubara at al. \cite{Kom74} \ in
uniaxial ferromagnet MnP (for longitudinal sound wave of \ 90 MHz
propagating along the $b$ axis) for the magnetic field applied along the
easy $c$ axis. The dashed lines correspond to the $T<T_{C}$ \ and the solid
lines represent the $T>T_{C}$ \ phase. For large and moderate magnetic
fields one can observe a characteristic region where the sound attenuation
in the high-temperature range \ exceeds \ the corresponding \ attenuation
value\ in \ the low-temperature phase (for the same reduced temperature $|t|$%
). As was shown earlier \ in \cite{Paw20} this characteristic region (called
a region II in \cite{Paw20}) is a manifestation of a shift of maximum of
attenuation towards the higher temperatures. What can also be easily
recognized from Fig. \ref{Fig3} is that \ the curves show much steeper slope
in the hydrodynamic regime \ ($\omega \tau \ll 1$)\ for $T>T_{C}$ than in
the low temperature phase. It follows from the fact that for the
high-temperature phase the relaxational term is proportional to $h^{2}$\ \
and is much larger than the fluctuation term. So the exponent \ \ for the
sound attenuation \ is equal to $\ \ x_{+}^{\mathrm{eff}}=z^{\mathrm{eff}%
}\nu ^{\mathrm{eff}}+\alpha ^{\mathrm{eff}}+2\Delta ^{\mathrm{eff}}$ \ \
where \ $\Delta ^{\mathrm{eff}}=\beta ^{\mathrm{eff}}+\gamma ^{\mathrm{eff}%
}. $ \ Here the effective exponents \cite{Paw09,Riedel74} which depend on
the temperature and magnetic field has been used. For example the
order-parameter exponent $\beta \simeq 0.326$ in the critical region but \ $%
\beta ^{\mathrm{eff}}=1/2$ in MFA regime or $\beta ^{\mathrm{eff}}=1/4$ in
the tricritical range. As was shown by Kuz'min \cite{Kuz08} \ in many
ferromagnets the equation of state can be described quite well by relatively
simple Landau Ginzburg theory with the functional (\ref{hop}). This means
that the critical fluctuations are confined to much narrower interval around
the critical temperature than was thought before and the single critical
exponent $\beta $ should be rather replaced by the effective exponent in
order to take account of the crossover\ from the asymptotic critical regime
to the Landau theory region. It follows from that analysis that the
proximity to the tricritical point \ should be also taken into account. It
was \ confirmed recently \ \cite{Paw20} that away from the critical point
the sound attenuation in MnP can be well described \ by the Landau Ginzburg
theory\ with the sixth-order term in the free energy playing the crucial
role. The crossover model \ presented here \cite{Paw22} has an advantage on
the previous approach \ that it satisfactorily describes the ultrasonic
attenuation coefficient also in critical regime i.e. for small magnetic
field and the reduced temperature. It should be noted that the Landau
Ginzburg theory gives $\ \alpha (h=0,T>T_{c})=0$ and $\alpha (h=0,T<T_{c})%
\overset{T\rightarrow T_{c}}{\longrightarrow }0$ , which is contrary to the
experiment. From Fig. \ref{Fig3} one can observe that the displacement of
the sound attenuation peak \ is monotonic with respect to the intensity of
the magnetic field. \ The maximum moves towards higher temperatures. \

\begin{figure}[h]
\begin{center}
\begin{minipage}{0.32\columnwidth}
        \centering\includegraphics[width=.95\columnwidth]{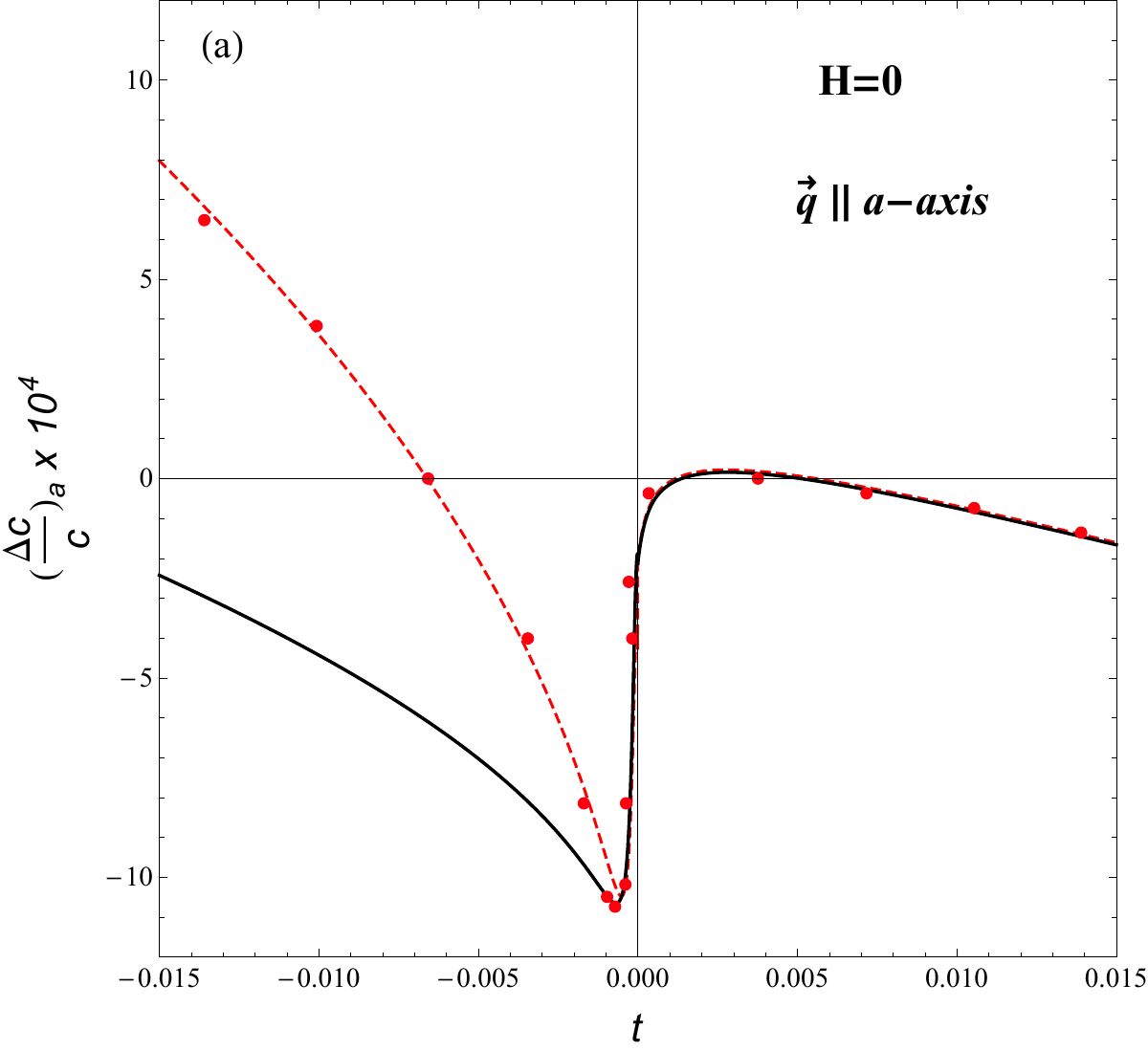}
  \end{minipage}
\begin{minipage}{0.32\columnwidth}
        \centering\includegraphics[width=.95\columnwidth]{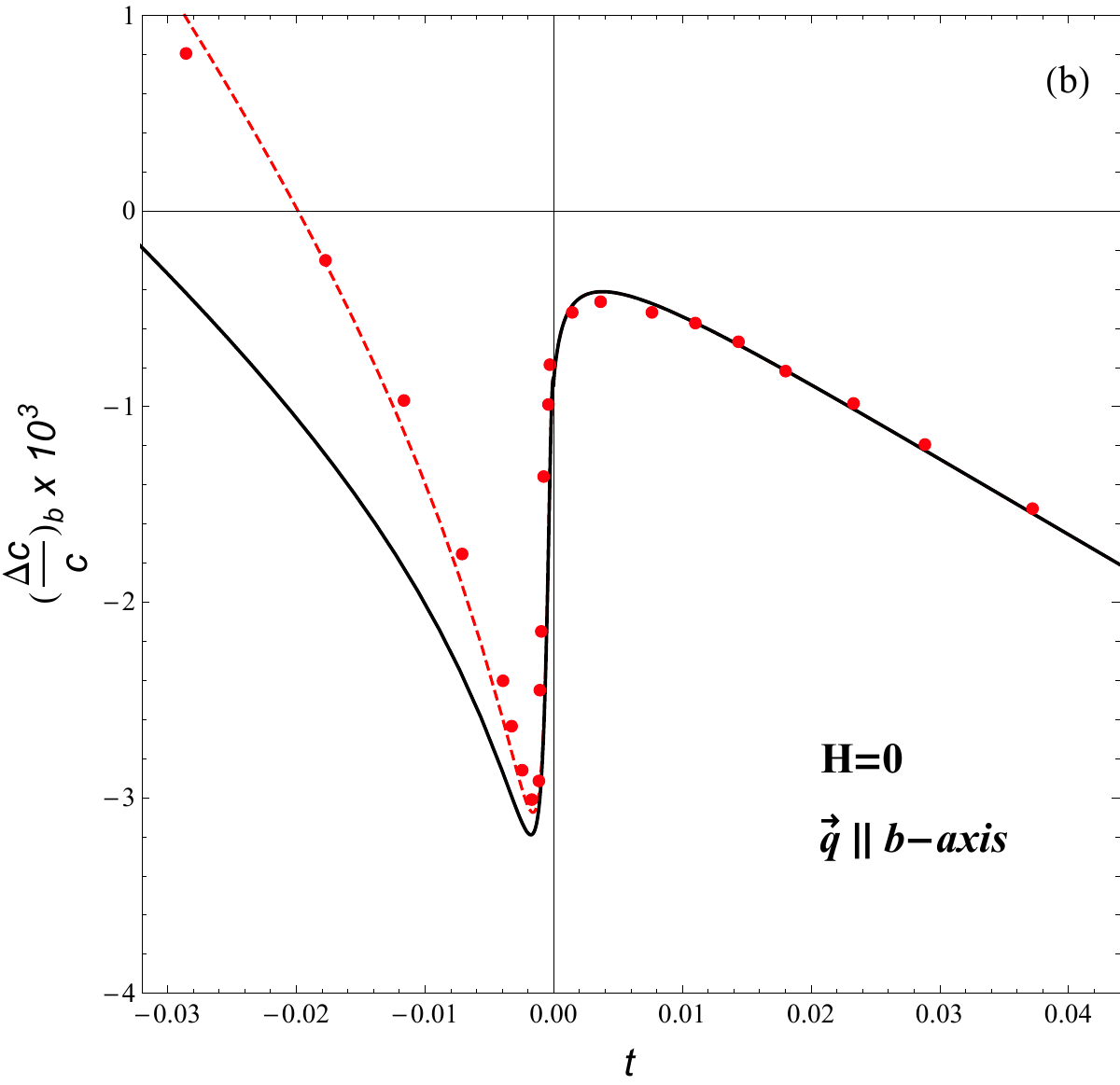}
  \end{minipage}
\begin{minipage}{0.32\columnwidth}
        \centering\includegraphics[width=.95\columnwidth]{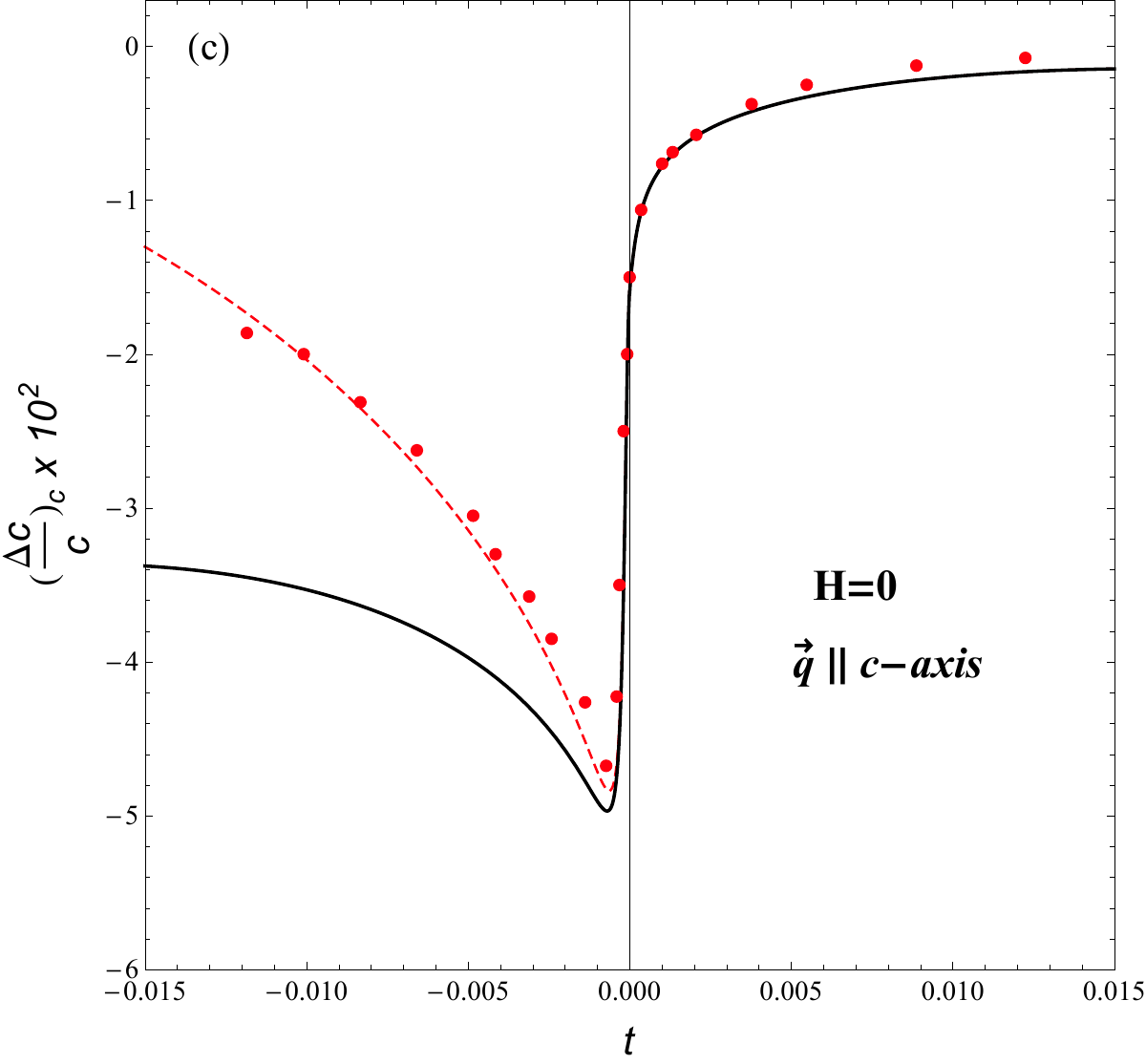}
  \end{minipage}
\end{center}
\par
\vspace{-0.5cm}
\caption{Sound velocity change in MnP versus the reduced temperature $t$.
The continuous lines correspond to the parametric model and the dashed lines
contain also phenomenological term proportional to the square of
magnetization in order to improve the fit. a) 10 MHz longitudinal wave along
the $a$ axis \protect\cite{Ferry}. b) 30 MHz longitudinal wave propagating
along the $b$ axis \protect\cite{Ishi77}. c) 10 MHz longitudinal wave
propagating along the $c$ axis \protect\cite{Ferry}. }
\label{Fig4}
\end{figure}
It would be interesting to see if \ the parameters obtained from fitting our
model to the attenuation data lead also to sensible predictions for the
sound wave velocity. In Fig. \ref{Fig4}\ \ \ we show the relative sound
velocity changes
\begin{equation}
\frac{\Delta c}{c_{0}}\equiv \frac{c(t,\omega )-c_{0}}{c_{0}}%
=Ac_{0}Y(r)^{-\alpha /\Delta _{s}}\func{Re}\left[ f_{\text{rel}}(y,\theta
,r)+\frac{A_{\text{fluc}}}{A_{\text{rel}}}f_{\text{fluc}}(y)\right]
\label{Dc}
\end{equation}%
calculated for our model and compared with experimental data obtained by
Ishizaki et al. \cite{Ishi77} and by Ferry and Golding \cite{Ferry}.
It is known that in MnP the easy, intermediate and hard axes of
magnetization in the ferromagnetic phase are the $c$, \ $\ b$ $\ $and $a$\
axes, respectively. For the 30 MHz\ longitudinal sound wave propagating
along $b$-axis (Fig. \ref{Fig4}b) we have used the same coupling constant $%
A_{(b)}$ \ as for the sound attenuation coefficient predictions presented in
Fig. \ref{Fig3}, but of course it is obvious that the magnetostrictive
coupling is weakest in MnP along $\ $the $a$ axis \cite{Ferry} and strongest
along $\ $the $c$ axis \ so \ different coupling constants have to be used
for all three directions of propagation. In Fig. \ref{Fig4} we \ compare our
predictions\ with \ the measurements of$\ \ $velocity changes $\ \Delta
c/c_{0}$ for 10 MHz longitudinal waves propagating along the $a$ \ and $c$
axes (Fig. \ref{Fig4}a and \ref{Fig4}c)\ \ for zero field \cite{Ferry}. \ It
is generally observed that the depth of \ the velocity dip, the location of
the extremum and the shape of the curve in the high-temperature phase is
correct and only the width of the theoretical curve in the ordered phase is
too wide \ for all theoretical curves. To improve the fitting we added to
the critical velocity a phenomenological term proportional to the square of
magnetization (the dashed curves). Such terms may appear in ferromagnets as
a result of higher order magnetoelastic couplings like $\varepsilon
^{2}S^{2} $ \ ($\varepsilon $ is the elastic strain) \ in the
Landau-Ginzburg interaction functional \cite{lut04} , or from other
mechanisms \cite{Pov98,Rou81}.
\begin{figure}[h]
\begin{center}
\begin{minipage}{0.6\columnwidth}
        \centering\includegraphics[width=.95\columnwidth]{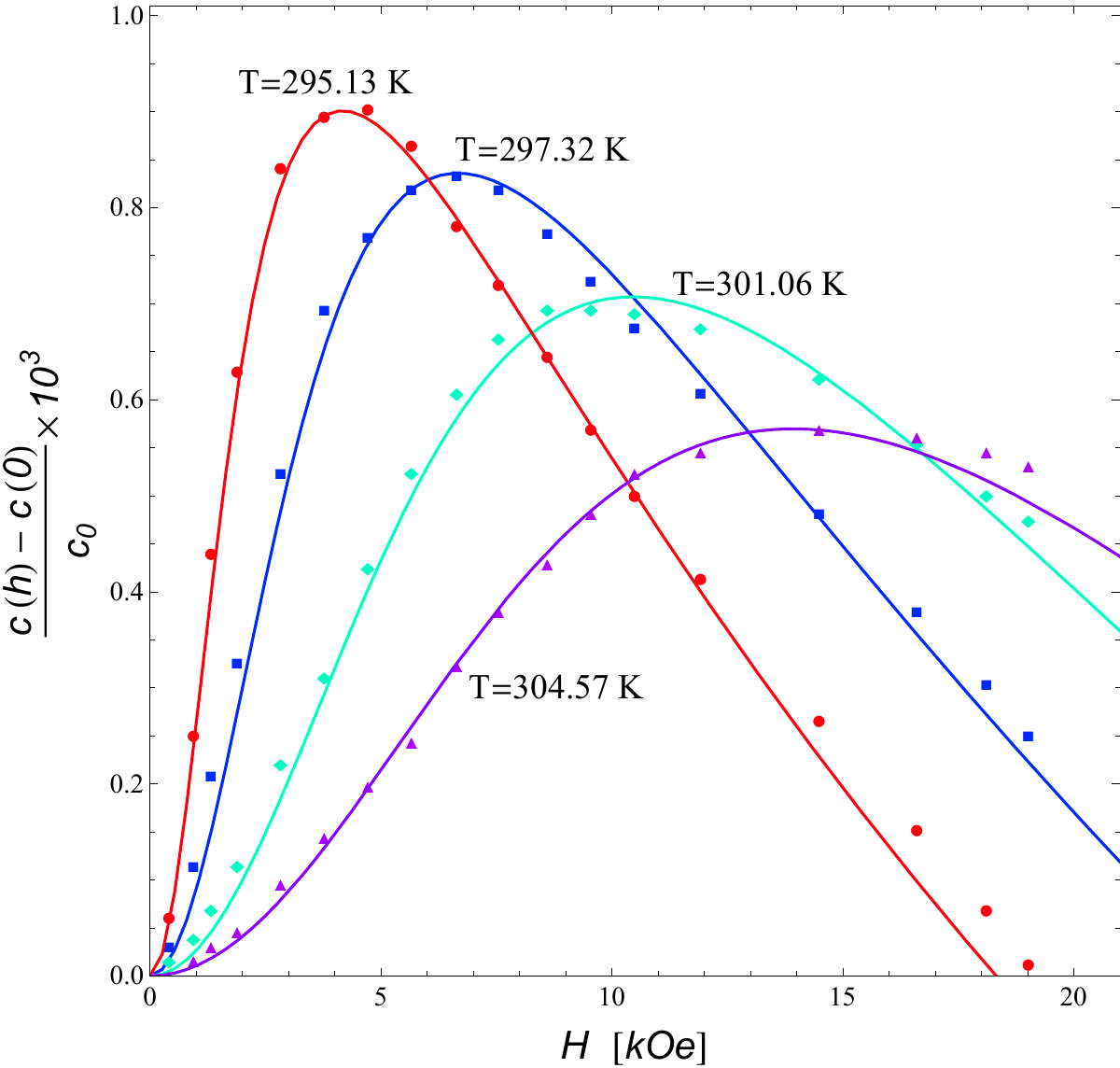}
  \end{minipage}
\end{center}
\par
\vspace{-0.5cm}
\caption{The magnetic field dependence of the sound velocity change in MnP
for several temperatures. The points show the measurements for the 30 MHz
longitudinal sound wave propagating along the $b$ axis \protect\cite{Ishi77}
with magnetic field applied along $c$ axis. }
\label{Fig5}
\end{figure}
In Fig. \ref{Fig5} we show the magnetic field behavior \ of the sound
velocity change
\begin{equation}
\frac{\Delta c(h)}{c_{0}}\equiv \frac{c(t,h)-c(t,0)}{c_{0}},
\end{equation}%
as a function of the magnetic field. For the longitudinal wave of 30 MHz \
propagating along the $b$ axis we took the same fitting parameters as in
Fig. \ref{Fig3}.\ The velocity change shows a quadratic increase in low
fields and decreases in high fields after showing a broad peak. Similar
behavior was reported \ for $\ \Delta \alpha (h)$ \ in our earlier work \cite%
{Paw20}. \

\begin{figure}[h]
\begin{center}
\begin{minipage}{0.6\columnwidth}
        \centering\includegraphics[width=.9\columnwidth]{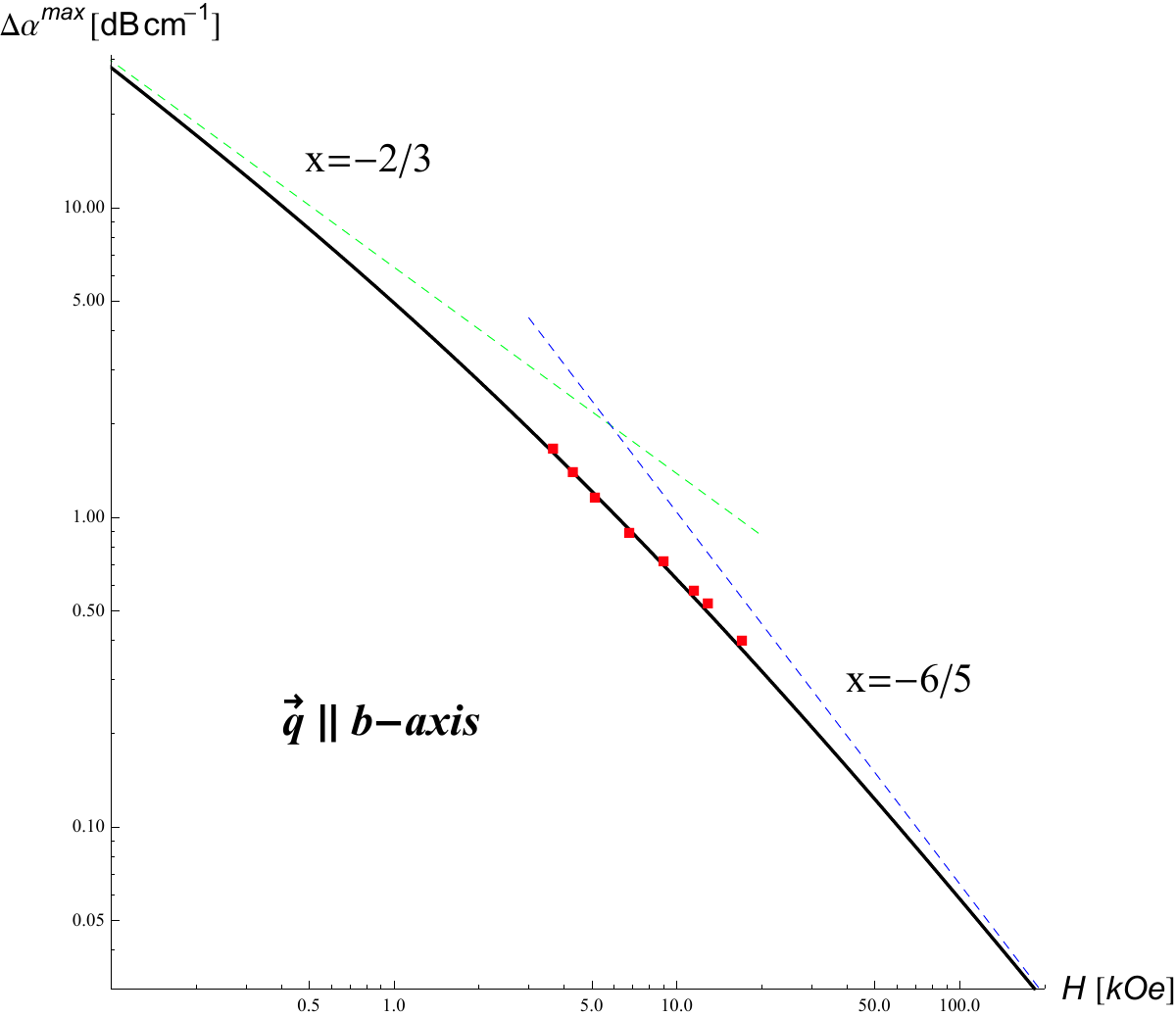}
  \end{minipage}
\end{center}
\par
\vspace{-0.5cm}
\caption{The magnetic field dependence of the sound attenuation peak height $%
\Delta \protect\alpha ^{\text{max}}$. The continuous line is an estimation
from Landau-Ginzburg theory, Eq. (\protect\ref{mfestim}), and the dashed
lines are the asymptotic power-law behavior $\Delta \protect\alpha ^{\text{%
max}}\propto h^{-2/3}$ correct for $Q=0$ or $\Delta \protect\alpha ^{\text{%
max}}\propto h^{-6/5}$ correct for $Q=\infty $. The points show the results
in MnP crystal for 90 MHz longitudinal sound wave propagating along the $b$
axis \protect\cite{Ishi77}. The magnetic field is directed along the easy $c$%
-axis. }
\label{Fig33}
\end{figure}

We present the height of the ultrasonic attenuation peak observed in the
field dependence of the sound attenuation in Fig. \ref{Fig33}, where the
Landau-Ginzburg estimation of this height is also shown. From Eq. (\ref%
{alfaMFA}) it follows that\ at large reduced temperatures,
\begin{equation}
\alpha (\omega ,t,h)\simeq \frac{2f_{0}^{2}\omega ^{2}}{c_{0}^{3}\Gamma }%
M^{2}(t,h)\chi _{GL}(t,h)^{2},  \label{mfestim}
\end{equation}%
because the reduced frequency $y$ is very small.\ \ Differentiating $%
M^{2}\chi _{GL}^{2}$\ \ \ with respect to $h$ \ we find the extremum
condition:
\begin{equation}
at^{\text{max}}=3uM_{\text{max}}^{2}+15vM_{\text{max}}^{5},  \label{maxcon}
\end{equation}%
and inserting it into the equation of state \ \ (\ref{ESMF}) \ we obtain
\begin{equation}
4uM_{\text{max}}^{2}+16vM_{\text{max}}^{5}=h/M_{\text{max}}.  \label{max2con}
\end{equation}%
The last equation gives us $M_{\text{max}}(h)$ and setting it into Eq. (\ref%
{mfestim}) we obtain the function $\Delta \alpha ^{\text{max}}(h).$\ \
Simple, power-law expressions are obtained for the characteristic quotient $%
Q $\ equal \ to $\infty $ and $0$. \ Thus, \ for the tricritical regime ($%
u=0)$ \ $\Delta \alpha ^{\text{max}}(h)\propto h^{-6/5}$ as well as \ $%
\Delta \alpha ^{\text{max}}(h)\propto h^{-2/3}$ for the Landau theory ($v=0$%
). In Fig. \ \ref{Fig33} the Landau-Ginzburg estimation of $\Delta \alpha ^{%
\text{max}}(h)$\ \ is compared with experimental data obtained by Ishizaki
at al. \cite{Ishi77} for the ultrasonic 90 MHz wave propagating along $b$%
-axis in MnP. \ The plot indicates that neither of the asymptotic regimes
can be associated with the experimental behavior \ in MnP in this narrow
range of observations.

\begin{figure}[h]
\begin{center}
\begin{minipage}{0.6\columnwidth}
        \centering\includegraphics[width=.9\columnwidth]{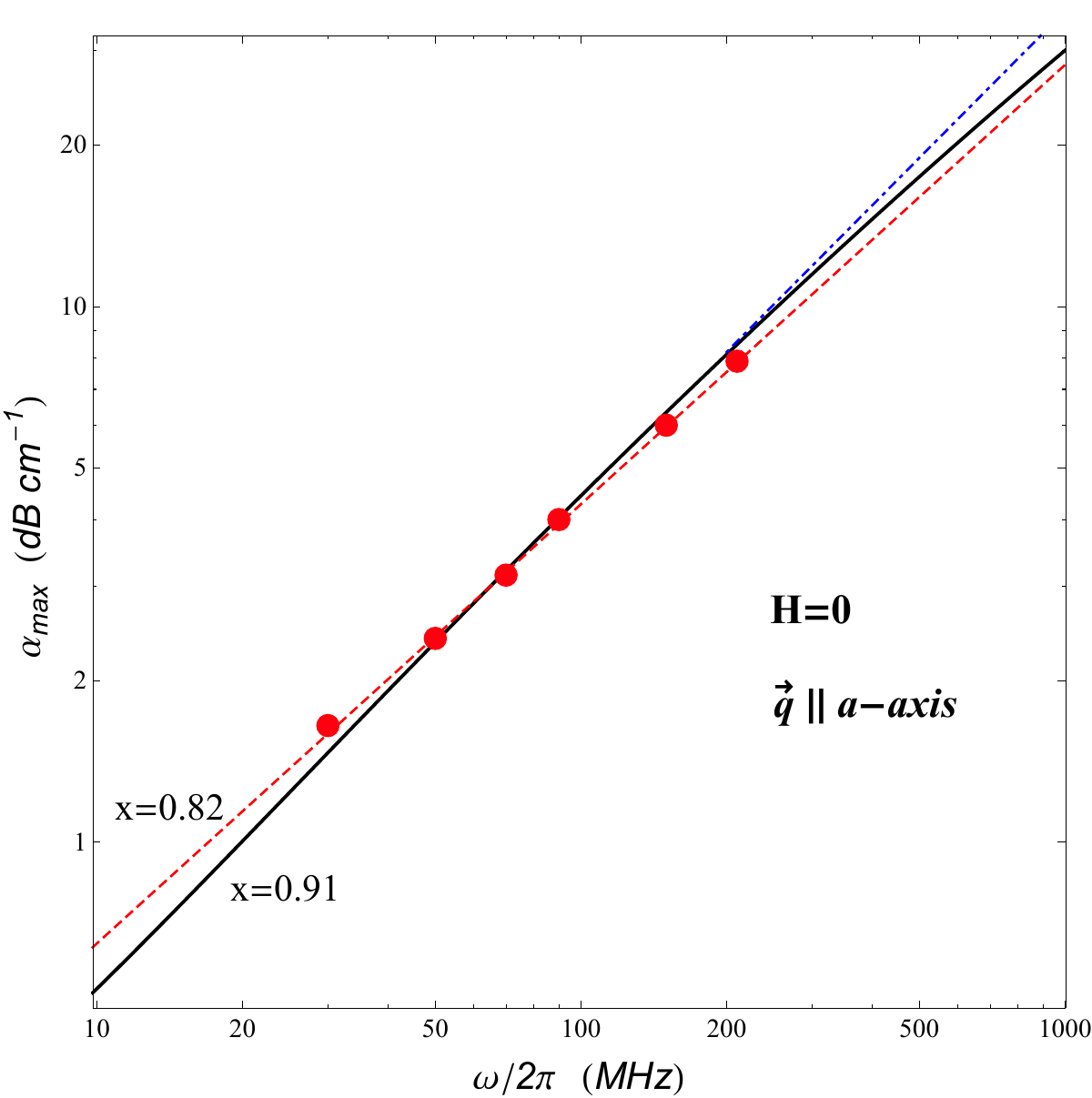}
  \end{minipage}
\end{center}
\par
\vspace{-0.5cm}
\caption{The frequency dependence of the sound attenuation at $T_{\mathrm{max%
}}$. The straight lines are power-law fits of the data (dashed line with
exponent $x=0.82$ \ and dot-dashed line with $x=0.91$) . Solid line
corresponds to Eq. (\protect\ref{alpar}) . The points show the measurements
for the longitudinal sound wave propagating along the $a$ axis \protect\cite%
{Suz82} . }
\label{Fig6}
\end{figure}
It would be of \ interest to see the dependence of the sound attenuation on
frequency. We consider the height of the ultrasonic attenuation peak
observed in the temperature dependence of the attenuation for vanishing
magnetic field. \ The critical attenuation shows \ a maximum at $T_{\text{max%
}}$, below the critical temperature, which shifts to lower temperature with
increasing frequency. Such measurements were performed \ in MnP \ by Suzuki
and Komatsubara \cite{Suz82} \ for the longitudinal waves propagating along $%
a$ axis.\ Comparing the height of the attenuation peaks for both propagation
directions at 90 MHz \cite{Kom74,Suz82} we estimated the ratio of the
critical attenuation amplitudes as $A_{(a)}/A_{(b)}\simeq 4/15$. This value
was also applied in Fig. \ref{Fig4}a and using this value \ \ we calculated
the heights of the sound attenuation maxima for several frequencies and
compared it with the experimental data from\ \cite{Suz82}. In Fig. \ref{Fig6}
\ it is seen that in the considered range of frequencies\ \ $\alpha _{\max
}(\omega )$ shows the power law behavior \ and our fit (the solid line) is
only slightly worse than the original experimental fit from the work of
Suzuki and Komatsubara \ \cite{Suz82} (dashed line), where the relation $%
\alpha _{\max }(\omega )\propto \omega ^{x}$ with $x=0.82$ \ was used. \ \ \
In our theory, the low-frequency asymptotic behavior, $\alpha _{\max
}(\omega )\propto \omega ^{1-\alpha /z\nu },$ is characterized by the
exponent $\ x=1-\alpha /z\nu \simeq 0.91.$

\section{Summary}

In this article we have performed calculations of the sound attenuation
coefficient and velocity within the parametric model containing sixth order
term. This model is valid not only the near the critical point but also
accounts for the crossover to simple Landau-Khalatnikov behavior \ or \ to
tricritical behavior. The constructed acoustic self-energy is valid in a
broad range of temperatures and magnetic fields. Crucial to the success of
this new approach is a feature that the MFA models \ usually lack the
ability to handle high value of the characteristic quotient $Q$ \ \cite%
{Kuz08}. \ Among other properties strongly affected by large value of the
characteristic quotient $Q$, one should mention the field-induced
displacement of the sound attenuation peak. \ It has been demonstrated that
such displacement is monotonic with respect to the intensity of the magnetic
field as well as with respect to the value of $Q$. \ Expressions obtained
for the sound attenuation coefficient and sound velocity are in fair
agreement with available experimental data \cite{Kom74,Ishi77,Ferry,Suz82}\
\ in manganese phosphide MnP.

\appendix

\section{Linear Parametric Representation}

In the asymptotic parametric representation the variables $r$ and \ $\theta $%
\ \ are defined in Eqs. (\ref{hh}-\ref{tt}) in the text and the scaled
equation of state in terms of these variables is Eq. (\ref{mm}), which,
despite its simplicity, gives \ a good experimental data approximation. Eq. (%
\ref{mm}) gives the attractively simple equation and the name of the linear
model obviously stems from it. \ For the linear model the singular Helmholtz
free energy can be obtained analytically \cite{SchLitHo69,HB72}\

\begin{equation}
A_{s}=h_{0}M_{0}r^{2-\alpha }f(\theta ),  \label{A1}
\end{equation}%
\begin{equation}
f(\theta )=f_{0}+f_{2}\theta ^{2}+f_{4}\theta ^{4},  \label{A2}
\end{equation}%
\begin{equation}
f_{0}=(1/2b^{4})[\delta -3-b^{2}\alpha (\delta -1)][(\delta +1)(\alpha
-1)\alpha ]^{-1},  \label{A3}
\end{equation}%
\begin{equation}
f_{2}=-(1/2b^{2})[\beta (\delta -3)-b^{2}\alpha (1-2\beta )][(\alpha
-1)\alpha ]^{-1},  \label{A4}
\end{equation}%
\begin{equation}
f_{4}=-\frac{1}{2}(1-2\beta )\alpha ^{-1}.  \label{A5}
\end{equation}%
\ \ The specific heat is given by

\begin{equation}
C_{h}-C_{B}=\left( \frac{h_{0}M_{0}}{2b^{4}(1-\alpha )\alpha }\right)
r^{-\alpha }c_{h}(\theta ),  \label{A6}
\end{equation}%
with
\begin{equation}
c_{h}(\theta )=\frac{(1-\alpha )l^{\prime }s-\beta \delta ls^{\prime }}{%
l^{\prime }k-\beta \delta lk^{\prime }},  \label{A7}
\end{equation}%
where $C_{B}$ represents an analytic fluctuation-induced background\
contribution \ \cite{Aga01} which is a smooth function of $\ t$ \ and the
functions $l(\theta )$ and $k(\theta )$ are defined in Eqs (\ref{hh}) and (%
\ref{tt}). The singular part of entropy satisfies:

\begin{equation}
S_{s}=\left( \frac{h_{0}M_{0}}{2b^{4}(1-\alpha )\alpha }\right) r^{1-\alpha
}s(\theta )
\end{equation}%
with the angular function $s(\theta )=\bar{s}_{0}+\bar{s}_{2}\theta ^{2}$
where
\begin{equation}
\bar{s}_{0}=\beta (\delta -3)-\alpha \beta (\delta -1)b^{2},  \label{A8}
\end{equation}%
\begin{equation}
\bar{s}_{2}=(\alpha -1)(\delta -3)\beta b^{2}.  \label{A9}
\end{equation}%
General expression for the susceptibility is%
\begin{equation}
\chi =\frac{M_{0}}{h_{0}}r^{-\gamma }X(\theta ),  \label{A10}
\end{equation}%
with
\begin{equation}
X(\theta )=\frac{k-\beta k^{\prime }\theta }{kl^{\prime }-\beta \delta
k^{\prime }l}.  \label{A11}
\end{equation}%
The critical exponents obey \ the usual scaling relations, $2-\alpha =\beta
(\delta +1)=\gamma +2\beta $ \cite{Stan72}. The coefficients $h_{0}$ and $%
M_{0}$ are system-dependent constants related to the critical amplitudes. \
\ \ \ \ \

%\appendix

\section{Crossover Parametric Representation}

The definition of the crossover model is:

\begin{equation}
h=h_{0}g^{\beta \delta -3/2}r^{3/2}\theta \lbrack Y^{(2\beta \delta
-3)/2\Delta _{s}}(1-\theta ^{2})+vr\theta ^{4}],  \label{B1}
\end{equation}

\begin{equation}
t=r(1-b^{2}\theta ^{2})\equiv rk(\theta ),  \label{B2}
\end{equation}

\begin{equation}
M=g^{\beta -1/2}M_{0}r^{1/2}Y^{\beta -1/2}\theta .  \label{B3}
\end{equation}%
The radial crossover function $Y$\ and its derivative are given by \cite%
{Aga01}%
\begin{equation}
1-(1-\bar{u})Y(r)=\bar{u}(1+\frac{\Lambda ^{2}}{arY(r)^{(2\nu -1)/\Delta
_{s}}})^{1/2}Y(r)^{\nu /\Delta _{s}},  \label{B4}
\end{equation}%
\begin{equation}
Y_{1}(r)=\frac{1}{\Delta _{s}}\frac{r}{Y}\frac{dY}{dr}=\frac{1}{\Delta _{s}}%
\frac{c_{1}\kappa ^{2}}{1+c_{1}c_{2}},  \label{B5}
\end{equation}%
with%
\begin{equation}
\kappa ^{2}=arY(r)^{(2\nu -1)/\Delta _{s}},  \label{B6}
\end{equation}

\begin{equation}
c_{1}=\frac{\Lambda ^{2}}{2\kappa ^{4}}\left( 1+\frac{\Lambda ^{2}}{\kappa
^{2}}\right) ^{-1}\left[ \frac{\nu }{\Delta _{s}}+\frac{(1-\bar{u})Y(r)}{%
1-(1-\bar{u})Y(r)}\right] ^{-1}  \label{B7}
\end{equation}

\begin{equation}
c_{2}=-(\frac{2\nu -1}{\Delta _{s}})\kappa ^{2}.  \label{B8}
\end{equation}

The thermodynamic functions are given by:

\begin{equation}
A_{s}(r,\theta )=\tilde{h}_{0}\tilde{M}_{0}r^{2}[Y^{-\alpha /\Delta
_{s}}f(\theta )+\frac{1}{6}vr\theta ^{6}],  \label{B9}
\end{equation}%
\begin{equation}
S=\tilde{h}_{0}\tilde{M}_{0}rY^{-\alpha /\Delta _{s}}\tilde{s}(\theta
,r)-C_{B}rk(\theta ),  \label{B10}
\end{equation}

\begin{equation}
C_{h}=\tilde{h}_{0}\tilde{M}_{0}Y^{-\alpha /\Delta _{s}}\widetilde{C}(\theta
,r)-C_{B},  \label{B11}
\end{equation}

\begin{equation}
\tilde{s}(\theta ,r)=-\frac{(2-\alpha Y_{1})\tilde{f_{0}}+[1-(\alpha +2\beta
-1)Y_{1}\tilde{f_{2}}\theta ^{2}-(\alpha +4\beta -2)Y_{1}\tilde{f_{4}}\theta
^{4}}{1-b^{2}[1-(1+(2\beta -1)Y_{1})\theta ^{2}]},  \label{B12}
\end{equation}%
\begin{equation}
\widetilde{C}(\theta ,r)=\frac{[(1-\alpha Y_{1})\tilde{s}+r\frac{\partial
\tilde{s}}{\partial r}|_{\theta }](l^{\prime }+5vrY^{\frac{3-2\beta \delta }{%
2\Delta _{s}}}\theta ^{4})-[(3/2+(\beta \delta -3/2)Y_{1})l+\frac{5}{2}vrY^{%
\frac{3-2\beta \delta }{2\Delta _{s}}}\theta ^{5}]\tilde{s}^{\prime }}{%
k(l^{\prime }+5vrY^{\frac{3-2\beta \delta }{2\Delta _{s}}}\theta
^{4})-k^{\prime }[(3/2+(\beta \delta -3/2)Y_{1})l+\frac{5}{2}vrY^{\frac{%
3-2\beta \delta }{2\Delta _{s}}}\theta ^{5}]},  \label{B13}
\end{equation}%
\begin{equation}
\chi =\frac{\tilde{M}_{0}}{\tilde{h}_{0}}r^{-1}Y^{(1-\gamma )/\Delta
_{s}}X(\theta ,r),  \label{B14}
\end{equation}%
\begin{equation}
X(\theta ,r)=\frac{k-[1/2+(\beta -1/2)Y_{1}]k^{\prime }\theta }{k(l^{\prime
}+5vrY^{\frac{3-2\beta \delta }{2\Delta _{s}}}\theta ^{4})-k^{\prime
}[(3/2+(\beta \delta -3/2)Y_{1})l+\frac{5}{2}vrY^{\frac{3-2\beta \delta }{%
2\Delta _{s}}}\theta ^{5}]}.  \label{B15}
\end{equation}

\end{document}